\documentclass[peerreview]{IEEEtran}

\usepackage[square,numbers,sort&compress]{natbib}
\usepackage{amsthm}
\usepackage{amsmath}
\usepackage{mathrsfs}
\usepackage{amssymb} 
\usepackage{bbm,dsfont} 
\usepackage{multirow} 
\usepackage{units}
\usepackage{stmaryrd}   
\usepackage{verbatim}
\usepackage{enumerate} 
\usepackage[table]{xcolor}
\usepackage{tikz}
\usetikzlibrary{calc}

\usepackage{xcolor}

\usepackage{hyperref} %
\usepackage%
{hyperref} %
\hypersetup{
    colorlinks,
    linkcolor={blue!80!black},%
    citecolor={green!30!black},
    urlcolor={blue!80!black}
}

\newcommand{\bieee}{\begin{IEEEeqnarray}{rCl}}
\newcommand{\eieee}{\end{IEEEeqnarray}}
\newcommand{\prob}[1]{\Pr\left(#1\right)}
\newcommand{\given}{\mid}
\newcommand{\cprob}[2]{\Pr\left(#1\given #2\right)}

\renewcommand{\mathbbm}[1]{\text{\usefont{U}{bbm}{m}{n}#1}} 
\newcommand{\eps}{\varepsilon}

\newcommand{\norm}[1]{\left\lVert#1\right\rVert}
\newcommand{\trace}{\mathrm{Tr}}

\newcommand{\identity}{\mathbbm{1}}
\newcommand{\kb}[1]{ | #1 \rangle\langle #1 | } 

\newcommand{\ie}{\emph{i.e.} }
\newcommand{\eg}{\emph{e.g.} }
\newcommand{\etal}{\emph{et al.} }
\newcommand{\cf}{\emph{cf.} }

\newcommand{\tm}{\widetilde{m}}	
\newcommand{\tM}{\widetilde{M}}

\newcommand{\ha}{\hat{a}}
\newcommand{\hb}{\hat{b}}
\newcommand{\hc}{\hat{c}}

\newcommand{\he}{\hat{e}}

\newcommand{\hm}{\hat{m}}
\newcommand{\htm}{\hat{m}}
\newcommand{\hn}{\hat{n}}

\newcommand{\hM}{\widehat{M}}

\newcommand{\Aset}{\mathcal{A}}

\newcommand{\Dset}{\mathcal{D}}
\newcommand{\Fset}{\mathcal{F}}

\newcommand{\Gset}{\mathcal{G}}
\newcommand{\Hset}{\mathcal{H}}

\newcommand{\Kset}{\mathcal{K}}
\newcommand{\Lset}{\mathcal{L}}
\newcommand{\Mset}{\mathcal{M}}
\newcommand{\Pset}{\mathcal{P}}
\newcommand{\Uset}{\mathcal{U}}

\newcommand{\Wset}{\mathcal{W}}
\newcommand{\Xset}{\mathcal{X}}
\newcommand{\Yset}{\mathcal{Y}}
\newcommand{\Zset}{\mathcal{Z}}

\newcommand{\markovC}[1]{%
\begin{tikzpicture}[#1]%
\draw (0,0.3ex) -- (1ex,0.3ex);%
\draw (0.5ex,0.3ex) circle (0.2ex);
\draw[white] (0.2ex,0) -- (0.5ex,0);%
\end{tikzpicture}%
}
\newcommand{\Cbar}{\markovC{scale=2}}

\theoremstyle{remark}	\newtheorem{theorem}{Theorem}
\theoremstyle{remark}	\newtheorem{lemma}[theorem]{Lemma}
\theoremstyle{remark}	\newtheorem{corollary}[theorem]{Corollary}
\theoremstyle{remark}	
\theoremstyle{remark} \newtheorem{definition}{Definition}
\theoremstyle{remark} \newtheorem{remark}{Remark}
\theoremstyle{remark}

\newcommand{\tset}{\Aset^{\delta}}													

\newcommand{\channel}{\mathcal{N}}

\newcommand{\inR}{\mathcal{R}}

\newcommand{\opC}{\mathcal{C}}

\title{The Quantum Multiple-Access Channel with Cribbing Encoders}

\author{
		\vspace{0.1cm}
    \IEEEauthorblockN{Uzi Pereg\IEEEauthorrefmark{1}\IEEEauthorrefmark{2}, Christian Deppe\IEEEauthorrefmark{1}, and Holger Boche\IEEEauthorrefmark{3}\IEEEauthorrefmark{2}\IEEEauthorrefmark{4}} \\
		\vspace{0.25cm}
    \IEEEauthorblockA{\normalsize \IEEEauthorrefmark{1}Institute of Communication Engineering, Technical University of Munich \\
		\IEEEauthorrefmark{2}Munich Center for Quantum Science and Technology (MCQST)\\
    \IEEEauthorrefmark{3}Theoretical Information Technology, Technical University of Munich\\
		\IEEEauthorrefmark{4}Cyber Security in the Age of Large-Scale Adversaries Exzellenzcluster (CASA)\\
    Email: {\tt $\{$uzi.pereg,christian.deppe,boche$\}$@tum.de}}
}

\begin{document}
\maketitle

{}

\begin{abstract}
Communication over a quantum multiple-access channel (MAC) with  cribbing encoders is considered,
whereby Transmitter 2 performs a measurement on a system that is entangled with Transmitter 1.
Based on the no-cloning theorem, perfect  cribbing is impossible.
This leads to the introduction of a MAC model with noisy cribbing.
In the 
causal and non-causal cribbing scenarios, 
Transmitter 2 performs the measurement before 
the input of Transmitter 1 is sent through the channel. Hence, Transmitter 2's cribbing may inflict a ``state collapse" for Transmitter 1.
Achievable regions are derived for each setting. Furthermore, 
a regularized capacity characterization is established for robust cribbing, \ie when the cribbing system contains all the information of the channel input.
 Building on the analogy between the noisy cribbing model and the relay channel, a partial decode-forward region is derived for a quantum MAC with %
non-robust cribbing.
For the classical-quantum MAC with cribbing encoders, the capacity region is determined with perfect cribbing of the classical input, and a cutset region is derived for noisy cribbing. In the special case of a classical-quantum MAC with a deterministic cribbing channel, the inner and outer bounds coincide. %
\end{abstract}


\maketitle

\section{Introduction}
The multiple-access channel (MAC) is among the most fundamental and well-understood models in network communication and information theory \cite{Ahlswede:74p,LiXu:20p}.  
The MAC is also referred to as the uplink channel \cite{VaeziDingPoor:19b}, since it is interpreted 
in cellular communication as the link from the mobiles to the base station \cite{ShirvanimoghaddamDohlerJohnson:17m}, and
in the satellite-based Internet of Things (IoT),  from  ground devices to a satellite in space \cite{GHGCA:18p}.
 Furthermore,  in a wireless local area network (WLAN), the MAC represents  the channel from the terminals to the access point \cite[Section 3.2]{TerplanMorreale:18b}.
In general, the signals of different transmitters may interfere with one another. In particular, in sequential decoding, the receiver first decodes the message of one of the transmitters, while treating the other signals as  noise. Then, this estimation can be used in order to reduce the effective noise for the estimation of the next message. 
Following this interplay, if a cognitive transmitter has access to the signal of another transmitter, this knowledge can be exploited such that the receiver will decode the messages %
with less noise.
Such scenarios naturally arise in wireless systems of cognitive radios \cite{GJMS:09p,ChenChenMeng:14p} and the Internet of Things (IoT) \cite{LYGTG:19p}.
This motivates the study of the MAC with cribbing encoders, \ie the channel setting where one transmitter has access to the signal of another transmitter \cite{WillemsvanderMeulen:85p}.

Cooperation in quantum communication networks has been extensively studied in recent years, following both experimental progress and theoretical discoveries \cite{vanLoockAltBecherBensonBocheDeppe:20p,BFSDBFJ:20b}. 
Quantum MACs are considered in the literature in various settings. %
Winter \cite{Winter:01p}  derived a regularized characterization for the classical capacity region of the quantum MAC (see also \cite{Savov:12z}).
Hsieh \etal \cite{HsiehDevetakWinter:08p} and Shi \etal \cite{ShiHsiehGuhaZhangZhuang:21p} addressed the model where each transmitter shares entanglement resources with the receiver independently. 
Furthermore, Boche and N\"otzel \cite{BocheNoetzel:14p} addressed the cooperation setting of a classical-quantum %
MAC with conferencing encoders, where the encoders exchange messages between them in a constant rate (see also \cite{DiadamoBoche:19a}).  Remarkably, Leditzky \etal \cite{LeditzkyAlhejjiLevinSmith:20p} have
 shown that sharing entanglement between transmitters can strictly increase the achievable rates for a \emph{classical} MAC. The
channel construction in \cite{LeditzkyAlhejjiLevinSmith:20p} is based on  a pseudo-telepathy game \cite{BrassardBroadbentTapp:05p} where quantum strategies guarantee a certain win and outperform classical strategies. %
The authors of the present paper \cite{PeregDeppeBoche:21p2} have recently shown that the dual property does not hold, \ie entanglement between receivers does not increase achievable rates. %
Related settings of the quantum MAC involve transmission of quantum information \cite{Yard:05z,YardHaydenDevetak:08p}, error exponents \cite{HayashiCai:17a}, non-additivity effects  \cite{CzekajHorodecki:08a}, security \cite{AghaeeAkhbari:19c,BochJanssenSaeedianaeeni:20p,DasHorodeckiPisarczyk:21a,ChakrabortyNemaSen:21a},  and computation codes \cite{HayashiVazquezCastro:21a}.

The sixth generation of cellular networks (6G) is expected to  achieve gains in terms of latency, resilience, computation power, and trustworthiness in future communication systems, such as the tactile internet \cite{FettwisBoche:21m}, which not only transfer data but also control physical and virtual objects, by using quantum resources \cite{DangAminShihadaAlouini:20p}.
Cooperation between trusted hardware and software components in future communication systems has the potential to isolate untrusted components such that the attack surface of the communication system is substantially reduced \cite{FettwisBoche:21m,FettwisBoche:22p}. Quantum resources and cooperation
offer additional advantages, in terms of performance gains for communication tasks and reduction of attack surface, and  
are of great potential for 6G networks  \cite{TKWIBD:20p,FitzekBoche:21p2}.
It is therefore interesting to investigate cooperation for quantum MACs as a technique to achieve robust efficient protocols for future applications.

\begin{figure*}[tb]
\caption{The quantum multiple-access channel $\Mset_{A_1 A_2\rightarrow B}$ with cribbing at Encoder 2. 
}

\includegraphics[scale=0.6,trim={-4cm 7cm 10cm 6.5cm},clip]{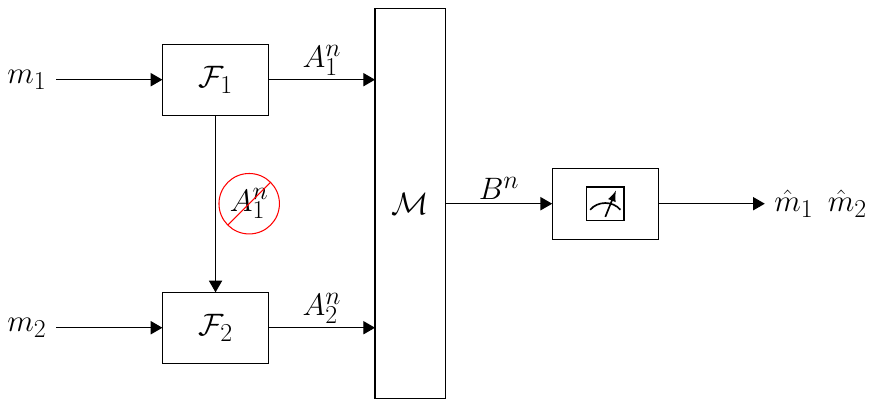} %
\\
{\footnotesize
(a) Perfect cribbing violates the laws of quantum mechanics. If Encoder 1 sends $A_1$ through the channel, then Encoder 2 is physically prohibited from having a copy of the input state.
}
\\ 
\includegraphics[scale=0.6,trim={-1cm 7cm 5cm 6.5cm},clip]{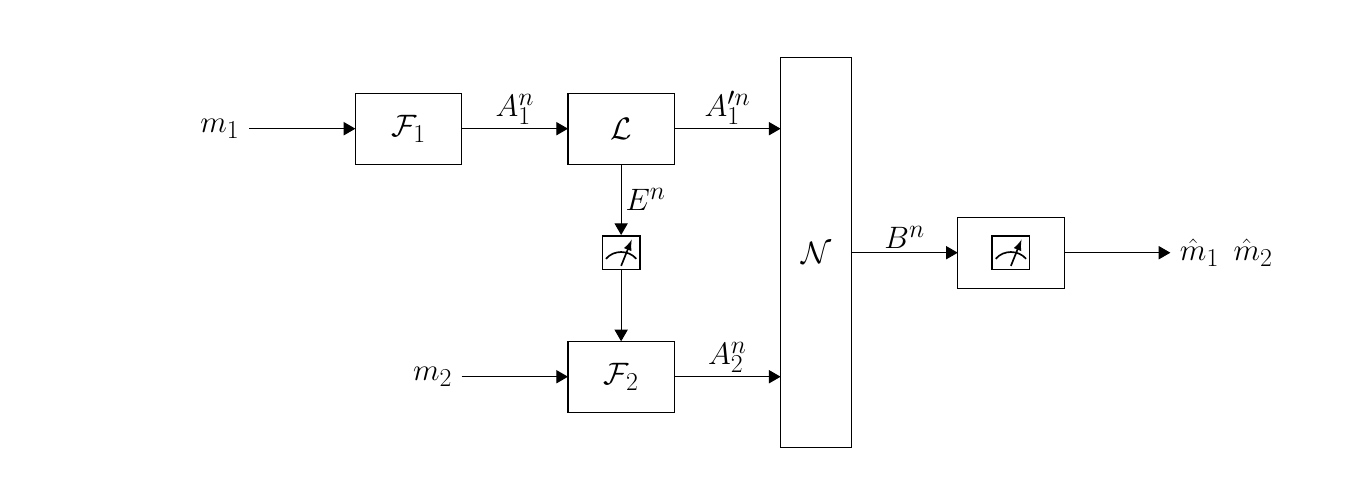} %
\\
{\footnotesize
(b) A quantum MAC $\channel\circ \Lset$ with noisy cribbing. Alice 1 sends the input system $A_1^n$ through $n$ copies of a cribbing channel $\Lset_{A_1\to A_1' E}$, producing two outputs, $A_1'^n$ and $E^n$.
Alice 2 performs a measurement on the system $E^n$, and uses the measurement outcome in order to encode the input state of $A_2^n$. Then, the input systems  $A_1'^n$ and $A_2^n$ are sent through $n$ copies of the communication channel $\channel_{A_1' A_2\to B}$.
The model can also be interpreted as if the second transmitter performs a measurement on the environment $E^n$ of the first transmitter.
}
\label{fig:Mcrib}
\end{figure*}

The classical MAC with cribbing encoders is a channel model with two transmitters, $X_1$ and $X_2$, and a single receiver, $Y$, where one transmitter 
has access to the other's transmissions. 
Willems  and van der Meulen \cite{WillemsVDMeulen:85p} 
introduced this setting and considered different scenarios of full cribbing, \ie with a perfect copy of the other sender's input. 
Suppose that Transmitter 2 knows $X_1$. 
Asnani and Permuter \cite{AsnaniPermuter:13p} pointed out that for a Gaussian channel, the full cribbing model is degenerate, as it reduces to full cooperation since a noiseless transmission of a continuous signal $X_1$ 
  allows  sending an infinite amount of information from Transmitter 1 to Transmitter 2.
This has motivated Asnani and Permuter to consider the MAC with partial cribbing, with a ``discretized" version of the other input. %
Specifically, Transmitter 2 may have access to $Z=g(X_1)$, instead of $X_1$, where $g(\cdot)$ is a deterministic function. See \cite{AsnaniPermuter:13p,KopetzPermuterShamai:16p} for further details.
The original work of Willems  and van der Meulen \cite{WillemsVDMeulen:85p} included different causality scenarios, where the
sender(s) have the $i$th copy of each other's inputs, either %
after sending their own transmission at time $i$ (strictly-causal cribbing), before the $i$th transmission (causal cribbing), or a priori before transmission takes place (non-causal cribbing). 
Classical cribbing is further studied in \cite{HelalBlochNosratinia:20p,HelalNostratinia:17c,%
HuleihelSteinberg:17p,HuleihelSteinberg:16c,Steinberg:14c,%
ZamanighomiEmadiChaharsooghi:11c,AmirSteinberg:12c,%
BracherLapidoth:14p,BrossLapidoth:10c%
}.

The MAC with cribbing encoders is closely-related to the relay channel \cite{ShimonovichSomekhBaruchShamai:13c,KolteOzgurPermuter:15c}. Even in the classical case, the capacity of the relay channel is an open problem.
Savov \etal \cite{SavovWildeVu:12c,Savov:12z} derived a partial decode-forward lower bound for the classical-quantum relay channel, where the relay encodes information in a strictly-causal manner. 
Recently, Ding \etal \cite{DingGharibyanHaydenWalter:20p} generalized those results and established the cutset, multihop, and coherent multihop bounds. %
Communication with the help of environment measurement can be modeled by a quantum channel with a classical relay in the environment \cite{HaydenKing:04a}.
Considering this setting, Smolin \etal \cite{Smolin:05p} and Winter \cite{Winter:05a} determined the environment-assisted quantum capacity
and classical capacity, %
respectively. %
Savov \etal \cite{SavovWildeVu:12c} further discussed future research directions of interest, and pointed out that quantum communication scenarios over the relay channel may have applications for the design of quantum repeaters \cite[Section V.]{SavovWildeVu:12c} (see also  \cite{DingGharibyanHaydenWalter:20p}).  In a recent work by the authors \cite{PeregDeppeBoche:21p2}, we have considered the quantum broadcast channel with conferencing receivers, %
and provided an information-theoretic perspective on quantum repeaters through this setting. %

In this paper, we consider the quantum MAC $\Mset_{A_1 A_2\to B}$ with cribbing encoders.
In quantum communication, the description is %
more delicate. 
By the no-cloning theorem, universal copying of quantum states is impossible.  Therefore, in the view of quantum mechanics, perfect  cribbing is against the laws of nature.
As illustrated in Figure~\ref{fig:Mcrib}, if Encoder 1 sends $A_1$ through the channel, then Encoder 2 is physically prohibited from having a copy of the input state. Hence, we consider the quantum MAC with noisy cribbing, consisting of a concatenation of a cribbing channel and the communication channel (see Figure~\ref{fig:Mcrib}b).
Specifically, Encoder 1 sends the input system $A_1$ through a cribbing channel that has two outputs, $A_1'$ and $E$.
Encoder 2 performs a measurement on the system $E$, and uses the measurement outcome in order to encode the input state of $A_2$. Then, the input systems  $A_1'$ and $A_2$ are sent through the communication channel.
The model can also be interpreted as if the second transmitter performs a measurement on the environment $E$ of the first transmitter.
The entanglement between the cribbing system $E$ and the communication channel input $A_1'$ has the following implication.
 If Encoder 2 measures $E_i$ before 
$A_{1,i}'$ is sent through the channel, as in the causal and non-causal scenarios, then Encoder 2's cribbing measurement may inflict a ``state collapse" of Encoder 1's transmission through the communication channel. In other words, in quantum communication, the cribbing operation interferes with Encoder 1's input before it is even transmitted through the communication channel.

We consider the scenarios of strictly-causal, causal, and non-causal cribbing.
For a MAC with \emph{robust cribbing},  the cribbing system $E$ includes all the information that is available in $A_1'$.
We derive achievable regions for each setting and establish
a regularized capacity characterization  for robust cribbing.
 Building on the analogy between the noisy cribbing model and the relay channel, we further develop a partial decode-forward region for the quantum MAC with strictly-causal non-robust cribbing.
For the classical-quantum MAC with cribbing encoders, the capacity region is determined with perfect cribbing of the classical input, and a cutset region is derived for noisy cribbing. In the special case of a classical-quantum MAC with a deterministic cribbing channel, the inner and outer bounds coincide.
The setting of noisy cribbing is significantly more challenging, as it is closely-related to the relay channel. The comparison between the relay channel and the MAC with cribbing encoders is further investigated in the present paper.

The paper is organized as follows. In Section~\ref{sec:def}, we give the definitions of the channel model and a brief review of related work. In Section~\ref{sec:Info}, we provide the information-theoretic tools for the analysis. 
The main results are given in Section~\ref{sec:main} and Section~\ref{sec:PDF}. In the former, we focus on robust cribbing, and in the latter, we introduce partial decode-forward coding for the non-robust case. The proofs are given in the appendix.

\section{Definitions and Related Work}
\label{sec:def}
\subsection{Notation, States, and Information Measures}
\label{subsec:notation}
 We use the following notation conventions. %
Script letters $\Xset,\Yset,\Zset,...$ are used for finite sets.
Lowercase letters $x,y,z,\ldots$  represent constants and values of classical random variables, and uppercase letters $X,Y,Z,\ldots$ represent classical random variables.  
 The distribution of a  random variable $X$ is specified by a probability mass function (pmf) 
	$p_X(x)$ over a finite set $\Xset$. %
 We use $x^j=(x_1,x_{2},\ldots,x_j)$ to denote  a sequence of letters from $\Xset$. %
 A random sequence $X^n$ and its distribution $p_{X^n}(x^n)$ are defined accordingly. 

The state of a quantum system $A$ is a density operator $\rho$ on the Hilbert space $\Hset_A$.
A density operator is an Hermitian, positive semidefinite operator, with unit trace, \ie 
 $\rho^\dagger=\rho$, $\rho\succeq 0$, and $\trace(\rho)=1$. The set of all density operators acting on $\Hset_A$ is denoted by $\mathscr{D}(\Hset_A)$. The state is said to be pure if $\rho=\kb{\psi}$, for some vector $|\psi\rangle\in\Hset_A$.
Define the quantum entropy of the density operator $\rho$ as
$%
H(\rho) \triangleq -\trace[ \rho\log(\rho) ]
$. %
Consider the state of a pair of systems $A$ and $B$ on the product space
$\Hset_A\otimes \Hset_B$.
Given a bipartite state $\sigma_{AB}$, %
define the quantum mutual information as
\begin{align}
I(A;B)_\sigma=H(\sigma_A)+H(\sigma_B)-H(\sigma_{AB}) \,. %
\end{align} 
Furthermore, conditional quantum entropy and mutual information are defined by
$H(A|B)_{\sigma}=H(\sigma_{AB})-H(\sigma_B)$ and
$I(A;B|C)_{\sigma}=H(A|C)_\sigma+H(B|C)_\sigma-H(AB|C)_\sigma$, respectively.
A quantum channel $\Pset_{A\to B}$ is a completely-positive trace-preserving (cptp) linear map from $\mathscr{D}(\Hset_A)$ to $\mathscr{D}(\Hset_B)$.
A measurement of a quantum system is specified in two equivalent manners. When the post-measurement state is irrelevant, we specify a measurement by a positive operator-valued measure (POVM), \ie a set of positive semi-denfinite operators  $\{ K_x \}_{x\in\Xset}$
such that $\sum_x K_x=\identity$.
According to the Born rule, if the system is in state $\rho$, then the probability to measure $x$ is  $p_X(x)=\trace(K_x \rho)$.
More generally, a measurement is defined  by a set of operators $\{W_x\}_{x\in\Xset}$ such that 
$\sum_x W_x^\dagger W_x=\identity$. %
If the system is in the state $\rho$, then the measurement outcome $X$ is distributed by $ p_X(x)=\trace(W_x^\dagger W_x \rho)$ for $x\in\Xset$. If $X=x$ was measured, then the post-measurement state is 
$\rho'\equiv (W_x \rho W_x^\dagger)/\trace(W_x^\dagger W_x \rho)%
$.
Furthermore, the quantum instrument $\Wset_{A\to\bar{A}X}$ of this measurement is  the linear map from the original state, before the measurement, to the joint state of the system after the measurement with the measurement outcome, \ie
\begin{align}
\Wset_{A\to\bar{A}X}(\rho)= \sum_{x\in\Xset} W_x \rho W_x^\dagger \otimes \kb{x} \,.
\end{align}

A quantum Markov chain is defined as follows. 
The quantum systems $A\Cbar B\Cbar C$ form a Markov chain if there exists a recovery channel $\Pset_{B\rightarrow BC}$ such that  $\rho_{ABC}=(\mathrm{id}_A\otimes \Pset_{B\rightarrow BC})(\rho_{AB})$ \cite{Sutter:18b}.
In general, this holds if and only if
\begin{align}
I(A;C|B)_\rho =0
\end{align}
(see \cite[Theorem 5.2]{Sutter:18b}).

\subsection{Quantum Multiple-Access Channel}
\label{subsec:Qchannel}
A quantum multiple-access channel (MAC) maps a quantum state at the senders' system to a quantum state at the receiver. 
Here, we consider a channel with two transmitters.
Formally, a quantum MAC  is a   cptp map 
$%
\Mset_{ A_1 A_2\rightarrow B}  %
$ %
corresponding to a quantum physical evolution.
We assume that the channel is memoryless. That is, if the systems $A_1^n=(A_{1,i})_{i=1}^n$ and $A_2^n=(A_{2,i})_{i=1}^n$ are sent through $n$ channel uses, then the joint input state $\rho_{ A_1^n A_2^n}$ undergoes the tensor product mapping
$%
\Mset_{ A_1^n A_2^n \rightarrow B^n}\equiv  \Mset_{ A_1 A_2\rightarrow B}^{\otimes n} %
$. %

We will consider a MAC with cribbing, where Transmitter 2 can measure a system $E$ that is entangled with Transmitter 1's system. 
Assume without loss of generality that the quantum MAC can be decomposed as
\begin{align}%
\Mset_{A_1 A_2\to B}(\rho_{A_1 A_2})=
\trace_E\left[  (\channel_{A_1' A_2\to B}\circ \Lset_{A_1\to A_1' E})(\rho_{A_1 A_2 }) \right] 
\label{eq:cribChannel}
\end{align}%
for all $\rho_{A_1 A_2}$, and some channel $\Lset_{A_1\to A_1' E}$.  In general, there always exists such a channel, since we can define $\Lset_{A_1\to A_1' E}$ such that the outputs $A_1'$ and $E$ are in a product state and $\rho_{A_1'}=\rho_{A_1}$.  However, the interesting case is when the transmitted system $A_1'$  and the cribbing system $E$ are correlated. 

 The transmitters and the receiver are often called Alice 1, Alice 2, and Bob. 
In the cribbing setting, Alice 1 first sends the input state of  $A_1^n$ through the channel $\Lset_{A_1^n\to A_1'^n E^n}\equiv \Lset^{\otimes n}_{A_1\to A_1' E}$. 
 Alice 2 gains access to the system $E^n$, and performs a measurement. Then, she encodes the state of her input $A_2^n$ using the measurement outcome, and transmits it to Bob. Henceforth, we will refer to the channel $\Lset_{A_1\to A_1' E}$ as the cribbing channel, and to $E^n$ as the cribbing system.
We denote the quantum MAC with cribbing by $\channel_{A_1' A_2\to B}\circ \Lset_{A_1\to A_1' E}$.

In the sequel, we will also be interested in the following special cases.

\begin{definition}
\label{def:robust}
Let $\channel\circ \Lset$ be a quantum MAC with cribbing. Given an input state $\theta_{A_1 A_0}$, where $A_0$ is an arbitrary reference system, let $\rho_{A_1' E A_0}=\Lset_{A_1\to A_1' E}(\theta_{A_1 A_0})$ be the output of the cribbing channel.
We say that the quantum MAC is with robust cribbing if $A_0\Cbar E\Cbar A_1'$ form a quantum Markov chain. 

That is, for every input $\theta_{A_1 A_0}$, there exists a recovery channel $\mathcal{P}^\theta_{E \to A_1' E}$ such that
\begin{align}
\rho_{A_1' E A_0}=\mathcal{P}^\theta_{E \to A_1' E}(\rho_{E A_0}) \,.
\end{align}
 
\end{definition}

We give the following intuition. In the cribbing setting, Alice 2 may have a noisy copy of Alice 1's input. The condition above means that Alice 2's copy is ``at least as good" as the one which is transmitted through the channel $\channel$. That is, the channel input $A_1'$ does not contain more information than %
the cribbing system $E$, which Alice 2 measures.
Next, we define  the classical-quantum MAC %
with noiseless cribbing, which is a special case of a quantum MAC with robust cribbing.

\begin{definition}
\label{def:cqPerfect}
The classical-quantum MAC  with \emph{noiseless} cribbing is defined by $\channel\circ \Lset$ such that 
the inputs $A_1$, $A_2$ are classical, and the cribbing channel simply copies Alice 1's input,  \ie $A_1\equiv X_1$, $A_2\equiv X_2$, and $\Lset=\text{id}_{X_1\to X_1 X_1}$. Hence, the cribbing system is an exact copy of Alice 1's input, \ie
\begin{align}
&E\equiv A_1'\equiv A_1\equiv   X_1 \,.
\end{align}
We denote the classical-quantum MAC with noiseless cribbing by $\channel_{X_1 X_2\to B}\circ \text{id}_{X_1\to X_1 X_1}$.
\end{definition}

If the cribbing observation is noisy, then it may not satisfy the robustness property.
\begin{definition}
\label{def:cqNoisy}
The classical-quantum MAC  with a \emph{noisy} cribbing channel $Q_{Z|X_1}$ is defined %
such that 
the inputs $A_1$, $A_1'$, $A_2$ and the cribbing system $E$ are classical, \ie $A_1\equiv A_1'\equiv X_1$, $A_2\equiv X_2$, and $E\equiv Z$,  while the cribbing channel is specified by the classical noisy channel $Q_{Z|X_1}$.
We denote the classical-quantum MAC with noisy cribbing by $\channel_{X_1 X_2\to B}\circ Q_{Z|X_1}$.
\end{definition}

\begin{remark}
Another simple example of robust cribbing is the case where $A_1'$ does not store information at all, say
$|\Hset_{A_1'}|=1$. The resulting channel %
is a basic multi-hop link that concatenates two point-to-point channels. Specifically, Alice 1 sends the input $A_1$ to Alice 2 via $\Lset_{A_1\to E}$, Alice 2 measures $E$, encodes the input $A_2$, and send it to Bob over $\channel_{A_2\to B}$.
\end{remark}

\subsection{Coding with Cribbing}
\label{sec:Coding}
We consider different scenarios of cribbing. 
First, we define the setting of a MAC where Alice 2 obtains her measurements of the systems $E_1,E_2,\ldots$ in a causal manner. That is, at time $i$, Alice 2 can measure the \emph{past and present} systems $E_1,\ldots, E_i$.

\begin{definition} %
\label{def:ClcapacityE}
A $(2^{nR_1},2^{nR_2},%
n)$ classical  code for the quantum MAC $\channel_{A_1' A_2\rightarrow B}\circ \Lset_{A_1\to A_1' E}$ with
cribbing  consists of the following: 
\begin{itemize}
\item 
Two message sets  $[1:2^{nR_1}]$ and $ [1:2^{nR_2}]$, assuming $2^{nR_k}$ is an integer;
\item
	an encoding map $\Fset_1: [1:2^{nR_1}]\to \mathscr{D}(\Hset_{A_1}^{\otimes n})$ for Encoder 1;
\item
a sequence of cribbing POVMs $\Kset_i=\{K^{z}_{E^i}\,,\; z\in\Zset\}  $, for Encoder 2;
\item
a sequence of $n$ encoding maps $\Fset_{2,i}: \Zset^i\times [1:2^{nR_2}]\to \mathscr{D}(\Hset_{A_2}^{\otimes i})$, where $i\in [1:n]$, for Encoder 2; and
\item
a decoding POVM   $\Dset= \{D^{m_1,m_2}_{B^n}\}  $, where the measurement outcome is an index pair  $(m_1,m_2)$, with $m_k\in[1:2^{nR_k}]$ for $k=1,2$.
\end{itemize}
The sequence of encoding maps needs to be consistent in the sense that the states $\rho^{m_2,z^i}_{A_2^{i}}\equiv \Fset_{2,i}(z_1^i,m_2)$ satisfy $\trace_{A_{2,i+1}^n}(\rho^{m_2,z_1^n}_{A_2^{n}})=\rho^{m_2,z_1^i}_{A_2^{i}}$.
We denote the code by $(\Fset_1,\Fset_2,\Kset,\Dset)$.
\end{definition}

The communication scheme is depicted in Figure~\ref{fig:Mcrib}b.  
The sender Alice 1 has the input system  $A_1^n$, the sender Alice 2  has both $A_2^n, E^n$, and the receiver Bob has  $B^n$. 
Alice $k$ chooses a message $m_k$ according to a uniform distribution over $[1:2^{nR_k}]$, for $k=1,2$.
 To send the message $m_1\in [1:2^{nR_1}]$,
Alice 1 encodes the message by $\rho_{A_1^n}^{m_1}\equiv \Fset_{1}(m_1)$, and sends her transmission $A_1^n$ through $n$ uses of the cribbing channel. %
Her transmission induces the following state,
\begin{align}
\rho^{m_1}_{A_1'^n E^n}= \Lset^{\otimes n} %
( \rho_{A_1^n}^{m_1}  ) %
\end{align}
where $E^n$ is the cribbing system which will be measured by the second transmitter. 

At time $i$, Alice 2 measures the system $E^i$ using the measurement $\Kset_i$, and obtains a measurement outcome $z_{i}$. To send the message $m_2\in [1:2^{nR_2}]$, she prepares the input state $%
\Fset_{2,i}(m_2,z^i)$ in a causal manner, and sends her transmission. 
The average joint input state is
\begin{align}%
\rho^{m_1,m_2}_{A_{1}'^n A_{2}^n}
=
\sum_{z^n\in\Zset^n} \trace_{E^n} \left[  (\mathrm{id}\otimes K^{z^n})  
\rho^{m_1}_{A_{1}'^n E^n} \right] \otimes \Fset_{2,n}(m_2,z^n)
\end{align}%
with $K^{z^n}_{E^n}\equiv K_{E^n}^{z_n} K_{E^{n-1}}^{z_{n-1}} \cdots K_{E_1}^{z_{1}} $.

Bob receives the channel output systems $B^n$ in the following state,
\begin{align}
\rho^{m_1,m_2}_{ B^n }=\channel%
^{\otimes n} (\rho^{m_1,m_2}_{A_1'^n A_2^n}) \,.
\end{align}
He measures with the decoding POVM $\Dset$ and obtains 
 an estimate of the message pair $(\htm_1,\htm_2)\in [1:2^{nR_1}]\times [1:2^{nR_2}]$ from the measurement outcome.
The average probability of error of the code is
\begin{align}
&P_{e}^{(n)}(\Fset_1,\Fset_2,\Kset,\Dset)= %
\frac{1}{2^{n(R_1+R_2)}}\sum_{m_1=1}^{2^{nR_1}} \sum_{m_2=1}^{2^{nR_2}} \left(1-\trace[  
D^{m_1,m_2 }%
\rho^{m_1,m_2}_{ B^n} ] \right)\,.
\end{align}

A $(2^{nR_1},2^{nR_2},n,\eps)$ classical code satisfies 
$%
P_{e}^{(n)}(\Fset_1,\Fset_2,\Kset,\Dset)\leq\eps $. %
A rate pair $(R_1,R_2)$ is called achievable with  causal cribbing   if for every $\eps>0$ and sufficiently large $n$, there exists a 
$(2^{nR_1},2^{nR_2},n,\eps)$ code. 
 The classical capacity region $\opC_{\text{caus}}(\channel\circ\Lset)$ of the quantum MAC with causal cribbing  
is defined as the set of achievable pairs $(R_1,R_2)$, where the subscript `caus' indicates causal cribbing. %

We will also put a considerable focus on strictly-causal and non-causal cribbing. In the strictly-causal setting, Alice 2 can measure the cribbing system only after she has sent her transmission at each time instance.
That is, a code with strictly-causal cribbing  is defined in a similar manner, where Alice 2 first transmits her input $A_{2,i}$, and only then measures $E^i$. Therefore, her input state at time $i$ can only depend on the \emph{past} measurements $z^{i-1}$, and  the $i$th encoding map  has the form $\Fset_{2,i}: \Zset^{i-1}\times [1:2^{nR_2}]\to \mathscr{D}(\Hset_{A_2}^{\otimes i})$. We denote the  capacity region for this scenario by $\opC_{\text{s-c}}(\channel\circ\Lset)$, where the subscript 's-c' stands for strictly-causal cribbing.

With non-causal cribbing,  Alice 2 gains access to the entire sequence of systems $E^n$ a priori, \ie before the beginning of her transmission. Thus, she can perform a joint measurement at time $i=1$, with a single POVM $\Kset=\{K^{\ell}_{E^n}\}  $, before sending $A_2^n$. Thereby, she can prepare a joint state $\rho_{A_2^n}^{m_2,\ell}$ of any form. The  capacity region with non-causal cribbing is %
denoted by $\opC_{\text{n-c}}(\channel\circ\Lset)$. %

\begin{remark}
\label{rem:relay1}
The MAC with strictly-causal cribbing is closely related to the relay channel.
In particular, if we impose $R_2=0$, then Alice 2 can only use the cribbing channel to enhance the transmission of information from Alice 1 to Bob. That is, Alice 2 acts as a relay in this case.
 In the sequel, %
we will introduce %
methods for  quantum cribbing that are inspired by relay coding techniques.  
It should be noted, however, that as opposed to the relay channel, our cribbing channel (`sender-relay link') is not affected by the transmission of Alice 2 (the `relay').
Specifically, Savov \etal \cite{SavovWildeVu:12c,Savov:12z} considered a 
 relay channel 
$\Pset_{A_1 A_2\to E B}$, where the transmitter sends a classical input $A_1\equiv X_1$, the relay receives $E$ and transmits $A_2\equiv X_2$, and the destination receiver receives the quantum output $B$. 
Hence, the state of $E$ in the relay model is affected by both $A_1$ and $A_2$.  
Here, on the other hand, the cribbing channel acts only on $A_1$.
Nonetheless, it appears that the technical challenges carry over from the relay channel to the present model.
\end{remark}

\begin{remark}
\label{rem:inclusion}
Based on the definitions above, we have the following relation between the capacity regions of the MAC with strictly-causal, causal, and non-causal cribbing: $\opC_{\text{s-c}}(\channel\circ\Lset)\subseteq \opC_{\text{caus}}(\channel\circ\Lset)\subseteq \opC_{\text{n-c}}(\channel\circ\Lset)$. This follows because, if Alice 2 has causal access to $E^i$, or to the entire sequence of $E^n$, then she can choose to measure only $E^{i-1}$  at time $i$.
Hence, every code with strictly-causal cribbing can also be used given causal or non-causal cribbing.
Similarly, a causal scheme can be performed given non-causal cribbing.
\end{remark}

\subsection{Related Work}
We briefly review the capacity result for the quantum MAC without cribbing. 
A code without cribbing is defined in a similar manner, where Alice 2 is not allowed to measure the cribbing systems $E^n$.
In this case, the model is fully described by $\Mset_{A_1 A_2\to B}$, ignoring its components $\Lset_{A_1\to A_1' E}$ and $\channel_{A_1' A_2\to B}$ (see (\ref{eq:cribChannel}) and Figure~\ref{fig:Mcrib}b).
Denote the capacity region without cribbing by $\opC_{\text{none}}(\Mset)$. Define a rate region $\inR_{\text{none}}(\channel\circ\Lset)$ as follows,
\begin{align}%
\inR_{\text{none}}(\Mset)=
\bigcup_{ p_U p_{X_1 |U} p_{X_2 |U} \,,\; \theta_{A_1}^{x_1}\otimes \zeta_{A_2}^{x_2} } %
\left\{ \begin{array}{rl}
  (R_1,R_2) \,:\;
	R_1 &\leq I(X_1;B|X_2 U)_\omega  \\
  R_2   &\leq I(X_2;B|X_1 U)_\omega\\
	R_1+R_2 &\leq I(X_1 X_2;B|U)_\omega
	\end{array}
\right\}
\label{eq:inRnone}
\end{align}%
where the union is over the joint distributions of the auxiliary random variables $U,X_1,X_2$ and the product state collections $\{\theta_{A_1}^{x_1}\otimes \zeta_{A_2}^{x_2}\}$,
with
\begin{align}
&\omega_{U X_1 X_2 B}= \sum_{u,x_1,x_2} p_U(u)p_{X_1|U}(x_1|u) p_{X_2|U}(x_2|u) \kb{u} \nonumber\\& \otimes\kb{x_1}\otimes\kb{x_2}\otimes \Mset_{A_1 A_2 \to B}(\theta_{A_1}^{x_1}\otimes \zeta_{A_2}^{x_2}) 
\end{align}
and $|\Uset|=3$, $|\Xset_k|\leq 3(|\Hset_{A_k}|^2+1)$, for $k=1,2$.

\begin{theorem}[see \cite{Winter:01p,Savov:12z}]
\label{theo:cribNone}
The capacity region of the quantum MAC $\Mset_{A_1 A_2\to B}$ without cribbing satisfies
\begin{align}
 \mathcal{C}_{\text{none}}(\Mset)=\bigcup_{n=1}^\infty \frac{1}{n}\mathcal{R}_{\text{none}}(\Mset^{\otimes n}) \,.
\end{align}
\end{theorem}

\begin{remark}
The union over $p_U$ in the RHS of (\ref{eq:inRnone}) can be viewed as a convex hull operation, hence the  region $\mathcal{R}_{\text{none}}(\Mset)$ is convex.
This also has the interpretation of operational time-sharing. Suppose that $p_U$ is a type. By employing a sequence of $T$ codes consecutively, each corresponds to a rate pair $(R^{(u)}_1,R_2^{(u)})$ over a sub-block of length $n\cdot p_U(u)$ for $u\in\{1,\ldots,T \}$, one  achieves a rate pair
$(R_1,R_2)$ that corresponds to the convex combination of the rates, \ie $R_k= \sum_{u\in\Uset} p_U(u)R^{(u)}_k$ for  $k=1,2$. The time-sharing argument implies that the operational capacity region, with or without cribbing, is a convex set in general. 
\end{remark}

\section{Information Theoretic Tools}
\label{sec:Info}
To derive our results, we use the quantum version of the method of types properties and techniques. 
%

Standard method-of-types concepts are defined as in \cite{Wilde:17b,Pereg:22p}.  
We briefly introduce the notation and basic properties while the detailed definitions can be found in \cite[Appendix A]{Pereg:22p}.
In particular, given a density operator $\rho=\sum_x p_X(x)\kb{x}$ on the Hilbert space $\Hset_A$,  
 denote the strong $\delta$-typical set that is associated with $p_X$ by $\tset(p_X)$,  define the strong $\delta$-typical subspace as the vector space that is spanned by $\{ |x^n\rangle \,:\; x^n\in \tset(p_X)  \}$,
 and let
 $\Pi^{\delta}(\rho)$ be the projector onto this subspace.  
The following inequalities follow from well-known properties of strong $\delta$-typical sets \cite{NielsenChuang:02b}, %
\begin{align}
\trace( \Pi^\delta(\rho) \rho^{\otimes n} )\geq& 1-\eps  \label{eq:UnitT} \\
 2^{-n(H(\rho)+c\delta)} \Pi^\delta(\rho) \preceq& \,\Pi^\delta(\rho) \,\rho^{\otimes n}\, \Pi^\delta(\rho) \,
\preceq 2^{-n(H(\rho)-c\delta)}
\label{eq:rhonProjIneq}
\\
\trace( \Pi^\delta(\rho))\leq& 2^{n(H(\rho)+c\delta)} \label{eq:Pidim}
\end{align}
 where $c>0$ is a constant.
Furthermore, for $\sigma_B=\sum_x p_X(x)\sigma_B^x$, %
let $\Pi^{\delta}(\sigma_B|x^n)$ denote the projector corresponding to the conditional strong $\delta$-typical set %
given the sequence $x^n$.
Similarly \cite{Wilde:17b}, %
\begin{align}
\trace( \Pi^\delta(\sigma_B|x^n) \sigma_{B^n}^{x^ n} )&\geq 1-\eps'  \label{eq:UnitTCond} \\
 2^{-n(H(B|X')_\sigma+c'\delta)} \Pi^\delta(\sigma_B|x^n) &\preceq \,\Pi^\delta(\sigma_B|x^n) \,\sigma_{B^n}^{x^ n}\, \Pi^\delta(\sigma_B|x^n) \, \nonumber\\
\preceq 2^{-n(H(B|X')_{\sigma}-c'\delta)}& \Pi^\delta(\sigma_B|x^n)
\label{eq:rhonProjIneqCond}
\\
\trace( \Pi^\delta(\sigma_B|x^n))\leq& 2^{n(H(B|X)_\sigma+c'\delta)} \label{eq:PidimCond}
\end{align}
where $c'>0$ is a constant, $\sigma_{B^n}^{x^n}=\bigotimes_{i=1}^n \sigma_{B_i}^{x_i}$.
If $x^n\in\tset(p_X)$, then %
\begin{align}
\trace( \Pi^\delta(\sigma_B) \sigma_{B^n}^{x^n} )\geq& 1-\eps' 
\label{eq:UnitTCondB}
\end{align}
 as well (see \cite[Property 15.2.7]{Wilde:17b}).

The definition can be further generalized to
the case of $\sigma_B^x=\sum_z p_{Z|X}(z|x) \hat{\sigma}_B^{x,z}$. %
Given a fixed pair $(x^n,z^n)\in\Xset^n\times \Zset^n$, divide the index set $[1:n]$ into the subsets $I(a,b|x^n,z^n)=\{ i: (x_i,z_i)=(a,b)  \}$, for $(a,b)\in\Xset\times\Zset$.
The projector onto the conditional strong $\delta$-typical subspace given $(x^n,z^n)$ is 
\begin{align}
\Pi^\delta(\sigma_B|x^n,z^n)\equiv %
\bigotimes_{a\in\Xset} \bigotimes_{b\in\Zset}  \Pi^\delta_{B^{I(a,b|x^n,z^n)}} (\hat{\sigma}_B^{a,b})
 \,.
\end{align}
Whereas, given a fixed  $x^n\in\Xset^n$, we let %
$I_1(a|x^n)=\{ i: x_i=a  \}$, for $a\in\Xset$,
and we define the conditional $\delta_1$-typical 
projector %
as
\begin{align}
\Pi^{\delta_1}(\sigma_B|x^n)\equiv %
\bigotimes_{a\in\Xset}  \Pi^{\delta_1}_{B^{I_1(a|x^n)}} (\sigma_B^{a})
 \,.
\end{align}

\section{Main Results}
\label{sec:main}
We state our results on the quantum MAC $\channel_{A_1' A_2\rightarrow B}\circ \Lset_{A_1\to A_1' E}$ with cribbing.

\subsection{Strictly-Causal Cribbing}
We begin with the MAC  with strictly-causal cribbing, where Alice 2 transmits $A_{2,i}$ at time $i$, and then measures the cribbing system $E_i$ after the transmission. Hence, she only knows the past measurement outcomes $z^{i-1}$ at time $i$.
We derive an achievable region, and give a regularized characterization for the capacity region in the special case of robust cribbing.
As a consequence, we determine the capacity region of the classical-quantum MAC with noiseless cribbing. 
 Define
\begin{align}%
\mathcal{R}^{\text{DF}}_{\text{s-c}}(\channel\circ \Lset)=
\bigcup_{ p_U p_{X_1 |U} p_{X_2 |U} \,,\; \theta_{A_1}^{x_1}\otimes \zeta_{A_2}^{x_2} }
\left\{ \begin{array}{rl}
  (R_1,R_2) \,:\;
	R_1 &\leq I(X_1;E|U)_\omega  \\
  R_2   &\leq I(X_2;B|X_1,U)_\omega\\
	R_1+R_2 &\leq I(X_1 X_2;B)_\omega
	\end{array}
\right\}
\label{eq:inRsc}
\end{align}%
with
\begin{align}
&\omega_{U X_1 X_2 A_1' E A_2}=%
 \sum_{u,x_1,x_2} p_U(u)p_{X_1|U}(x_1|u) p_{X_2|U}(x_2|u) \kb{u} %
\otimes\kb{x_1}\otimes\kb{x_2}\otimes \Lset_{A_1 \to A_1' E}(\theta_{A_1}^{x_1})\otimes \zeta_{A_2}^{x_2} \,,
\\
&\omega_{U X_1 X_2 B}= \channel_{A_1' A_2\to B}(\omega_{U X_1 X_2 A_1' A_2}) \,.
\end{align}
The superscript `DF' stands for decode-forward coding, %
referring to the coding scheme that achieves this rate region.
Our terminology follows the analogy with the relay model (see Remark~\ref{rem:relay1}).

Before we state the capacity theorem, we give the following lemma. In principle, one may use the property below in order to compute the region $\mathcal{R}^{\text{DF}}_{\text{s-c}}(\channel\circ \Lset)$ for a given channel.
\begin{lemma}
\label{lemm:CardCl}
The union in (\ref{eq:inRsc}) is exhausted by %
auxiliary variables $U$, $X_1$, $X_2$ 
with $|\Uset|\leq |\Hset_{B}|^2+2$, 
$|\Xset_1|\leq (|\Hset_{A_1}|^2+2)(|\Hset_{B}|^4+2)$, and   $|\Xset_2|\leq (|\Hset_{A_2}|^2+1)(|\Hset_{B}|^2+2)$.
\end{lemma}
The proof of  is based on the Fenchel-Eggleston-Carath\'eodory lemma \cite{Eggleston:66p}, using similar arguments as 
in \cite{YardHaydenDevetak:08p}.
 The details are given in 
Appendix~\ref{app:CardCl}.

\begin{theorem}
\label{theo:cribSC}
Consider a quantum MAC $\channel_{A_1' A_2\to B}\circ \Lset_{A_1\to A_1' E}$. %
\begin{enumerate}[1)]
\item
The rate region $\mathcal{R}^{\text{DF}}_{\text{s-c}}(\channel\circ \Lset)$ is achievable for the quantum MAC  with strictly-causal cribbing, \ie
\begin{align}
 \mathcal{C}_{\text{s-c}}(\channel\circ\Lset)\supseteq \mathcal{R}^{\text{DF}}_{\text{s-c}}(\channel\circ \Lset) \,.
\end{align}
\item
Given robust cribbing, the capacity region satisfies
\begin{align}
 \mathcal{C}_{\text{s-c}}(\channel\circ\Lset)=\bigcup_{n=1}^\infty \frac{1}{n}\mathcal{R}^{\text{DF}}_{\text{s-c}}(\channel^{\otimes n}\circ \Lset^{\otimes n}) \,.
\end{align}
\end{enumerate}
\end{theorem}
The proof of Theorem~\ref{theo:cribSC} is given in Appendix~\ref{app:cribSC}.
To the best of our knowledge, the achievability result above is new even for a classical channel.
In the proof, we extend the block Markov coding scheme \cite{WillemsVDMeulen:85p}, where Alice 1 encodes two messages in each block in a sequential manner, such that one message is new and the other overlaps with the previous block.
Willems and van der Muelen \cite{WillemsVDMeulen:85p} refer to those messages as fresh information and resolution, respectively.
Given strictly-causal cribbing, Alice 2 can measure the cribbing systems of the previous block.
Thus, Alice 2 decodes the resolution by measuring the previous cribbing block, and encodes the resolution along with her own message. Bob recovers the messages in a reversed order using backward decoding.
We refer to this coding scheme as `decode-forward', since Alice 2 is responsible for decoding the messages of Alice 1, and forwarding them to Bob. 

\begin{remark}
\label{rem:robust}
The decode-forward coding scheme relies heavily on the ability of Alice 2 to decode using a cribbing measurement. 
Our results suggest that this is optimal when the cribbing link is robust. However, as we will discuss in Section~\ref{sec:PDF}, this is far from optimal when the cribbing link is too noisy. Therefore, we will introduce a new cribbing method which is  %
useful  for the non-robust case as well.   
\end{remark}

As a consequence of Theorem~\ref{theo:cribSC}, we determine the capacity region of the classical-quantum MAC with a noiseless cribbing channel (see Definition~\ref{def:cqPerfect}).
\begin{corollary}
\label{coro:cribSCcq}
The capacity region of the classical-quantum MAC $(\channel_{X_1 X_2\to B}\circ \text{id}_{X_1\to X_1 X_1})$ with strictly-causal noiseless cribbing is given by 
\begin{align}%
\mathcal{C}_{\text{s-c}}(\channel\circ\text{id})=%
\bigcup_{ p_U p_{X_1 |U} p_{X_2 |U}  }
\left\{ \begin{array}{rl}
  (R_1,R_2) \,:\;
	R_1 &\leq H(X_1|U)  \\
  R_2   &\leq I(X_2;B|X_1,U)_\omega\\
	R_1+R_2 &\leq I(X_1 X_2;B)_\omega
	\end{array}
\right\}
\label{eq:inRscCQ}
\end{align}%
with
\begin{align}
&\omega_{U X_1 X_2 B}= \sum_{u,x_1,x_2} p_U(u)p_{X_1|U}(x_1|u) p_{X_2|U}(x_2|u) \kb{u} %
\otimes\kb{x_1}\otimes\kb{x_2}\otimes
\channel_{X_1 X_2\to B}(x_1,x_2)
\end{align}
\end{corollary}
The proof of Corollary~\ref{coro:cribSCcq} is given in Appendix~\ref{app:cribSCcq}.
The corollary can be viewed as the classical-quantum counterpart of Willems and van der Muelen's result on the classical MAC with noiseless cribbing \cite{WillemsVDMeulen:85p}.

\subsection{Causal and Non-causal Cribbing}
In this section, we address causal and non-causal  cribbing. Recall that in the causal setting, Alice 2 measures the cribbing system $E_i$ at time $i$, before she transmits. Hence, she knows the past and present measurement outcomes $z^i=(z^{i-1},z_i)$ at time $i$. Whereas, in the non-causal case, Alice 2 can perform a joint measurement on $E^n$ a priori, \ie before the beginning of her transmission.
Here, as opposed to the model in the previous section, Alice 2's measurement may cause a ``state collapse" of Alice 1's input. Hence, it affects the input state for both transmitters. This can be seen in the achievable region below as well.
Define
\begin{align}%
\mathcal{R}^{\text{DF}}_{\text{caus}}(\channel\circ\Lset)=
\bigcup_{ p_U p_{X_1|U} \,,\; \Wset_{E\to \bar{E}Z} \,,\; p_{X_2|Z,U} \,,\; \theta_{A_1}^{x_1}\otimes \zeta_{A_2}^{x_2} }%
\left\{ \begin{array}{rl}
  (R_1,R_2) \,:\;
	R_1 &\leq I(X_1;\bar{E}Z|U)_\omega  \\
	R_2 &\leq I(X_2;B|X_1 U)_\omega  \\
	R_1+R_2 &\leq I(X_1 X_2;B)_\omega
	\end{array}
\right\}
\label{eq:inRcaus}
\end{align}%
where the union is over the probability distributions $p_U p_{X_1|U}$ for the auxiliary random variables $U$ and $X_1$, 
over the measurement instrument $\Wset_{E\to \bar{E}Z}(\rho)\equiv \sum_z W_z \rho W_z^{\dagger}\otimes \kb{z}$, the conditional probability distributions $p_{X_2|Z,U}$ for the auxiliary random variable $X_2$, and the input state collections $\{\theta_{A_1}^{x_1}\otimes \zeta_{A_2}^{x_2}\}$, with
\begin{align}
&\omega_{U X_1 A_1' \bar{E} Z X_2 A_2}= \sum_{u,x_1,z} p_U(u)p_{X_1|U}(x_1|u)  \kb{u}\otimes\kb{x_1}
\nonumber\\
&
\otimes W_z\left(\Lset_{A_1 \to A_1' E}(\theta_{A_1}^{x_1})\right)W_z^\dagger\otimes \kb{z}
\otimes 
\left(\sum_{x_2} p_{X_2|Z,U}(x_2|z,u) \kb{x_2}\otimes \zeta_{A_2}^{x_2}\right) \,,
\\
&\omega_{U X_1 X_2 B}= \channel_{A_1' A_2\to B}(\omega_{U X_1 X_2 A_1' A_2}) \,.
\end{align}
Here, $\Wset_{E\to \bar{E}Z}$ is a quantum instrument of a measurement, where $\bar{E}$ is the post-measurement cribbing system, and  $Z$ is  the measurement outcome.

\begin{theorem}
\label{theo:cribC}
Consider a quantum MAC $\channel_{A_1' A_2\to B}\circ \Lset_{A_1\to A_1' E}$. %
\begin{enumerate}[1)]
\item
The rate region $\mathcal{R}^{\text{DF}}_{\text{caus}}(\channel\circ \Lset)$ is achievable for the quantum MAC  with causal cribbing, \ie
\begin{align}
 \mathcal{C}_{\text{caus}}(\channel\circ\Lset)\supseteq \mathcal{R}^{\text{DF}}_{\text{caus}}(\channel\circ \Lset) \,.
\end{align}
\item
For the classical-quantum MAC $(\channel_{X_1 X_2\to B}\circ \text{id}_{X_1\to X_1 X_1})$ with noiseless cribbing,
\begin{align}%
\mathcal{C}_{\text{caus}}(\channel\circ\text{id})=\mathcal{C}_{\text{n-c}}(\channel\circ\text{id})=
\bigcup_{  p_{X_1  X_2 }  }
\left\{ \begin{array}{rl}
  (R_1,R_2) \,:\;
	R_1 &\leq H(X_1)  \\
  R_2   &\leq I(X_2;B|X_1)_\omega\\
	R_1+R_2 &\leq I(X_1 X_2;B)_\omega
	\end{array}
\right\}
\label{eq:inRcCQ}
\end{align}%
with
\begin{align}
&\omega_{ X_1 X_2 B}= \sum_{x_1,x_2} p_{X_1 X_2}(x_1,x_2)\kb{x_1,x_2}%
\otimes
\channel_{X_1 X_2\to B}(x_1,x_2)
\end{align}
\end{enumerate}
\end{theorem}
The proof of Theorem~\ref{theo:cribC} is given in Appendix~\ref{app:cribC}.
Part 1 seems to be new for classical channels as well, while part 2 is the classical-quantum version of Willems and van der Muelen's result  \cite{WillemsVDMeulen:85p}.

\begin{remark}
The classical achievability proof of Willems and van der Meulen \cite{WillemsvanderMeulen:85p} is based on the notion of Shannon strategies, as originally introduced in models of channel uncertainty \cite{KeshetSteinbergMerhav:07n}.
For the classical MAC, a strategy is defined as a function $f:\Xset_1\to\Xset_2$ that maps an input symbol of Alice 1 to that of Alice 2. 
Here, we replace the strategy by a quantum instrument that Alice 2 performs on the cribbing system.
\end{remark}

\vspace{-0.75cm}
\begin{center}
\begin{figure}[tb]
\center
\includegraphics[scale=0.52,trim={-0.5cm 11cm 9cm 11.5cm},clip]
{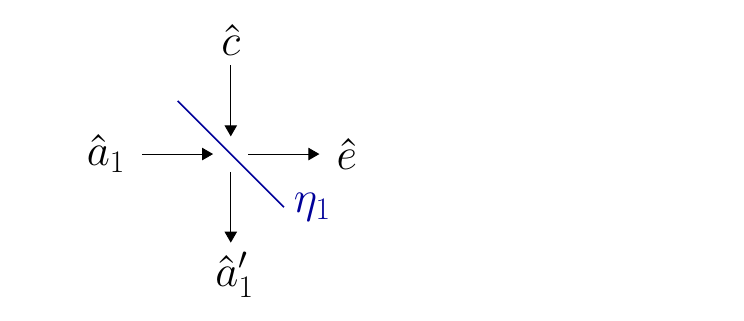} %
\hspace{-2cm}
\includegraphics[scale=0.52,trim={1cm 11cm 14cm 11.25cm},clip]
{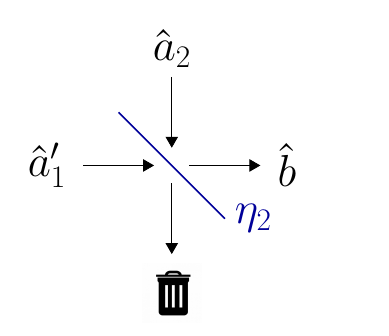} %
\caption{The beam splitter relation of the single-mode bosonic MAC. The left beam splitter corresponds to the cribbing channel $\Lset_{A_1\to A_1' E}$, while the right beam splitter describes the communication channel $\channel_{A_1' A_2\to B}$. 
Alice 1 encodes the message $m_1$ by a coherent state, and sends her transmission through $n$ uses of the cribbing channel $\Lset_{A_1\to A_1' E}$. %
The output $E^n$ is the cribbing system which will be measured by the second transmitter in a strictly-causal manner. 
At time $i$, Alice 2 encodes a coherent state that depends on her message $m_2$ and on the previous cribbing measurement outcomes $z^{i-1}$, she sends the state, and then performs a
 measurement on $E^i$ to obtain a measurement outcome $z_i$ which she will use in the next time instance. 
The inputs $A_{1,i}'^n$ and $A_{2,i}$ are sent through  the communication channel $\channel_{A_1' A_2\to B}$, and Bob receives the output $B_i$.  Bob performs a measurement on $B^n$ to obtain an estimate  $(\htm_1,\htm_2)$ of the senders' messages.
}
\label{fig:BSp}
\end{figure}
\end{center}

\subsection{Bosonic MAC with Strictly-Causal Cribbing}
To demonstrate our results, consider the single-mode bosonic MAC. We extend the finite-dimension result in Theorem~\ref{theo:cribSC} to the bosonic channel with infinite-dimension Hilbert spaces based on the discretization limiting argument by Guha \etal \cite{GuhaShapiroErkmen:07p}. A detailed description of (continuous-variable) bosonic systems can be found in \cite{WPGCRSL:12p}. Here, we only define the notation for the quantities that we use.
We use hat-notation, \eg $\ha$, $\hb$, $\he$, to denote operators that act on a quantum state.
The single-mode Hilbert space is spanned by the Fock basis 
$\{ |n\rangle \}_{n=0}^\infty$. Each $|n\rangle$ is an eigenstate of the number operator $\hn=\ha^\dagger \ha$, where $\ha$ is the bosonic field annihilation operator. In particular,  $
|0\rangle$ is the vacuum state of the field. 
The \emph{creation operator} $\ha^{\dagger}$ creates an excitation: 
$\ha^{\dagger}|n\rangle=\sqrt{n+1}|n+1\rangle$, for $n\geq 0$. Reversely, the \emph{annihilation operator} $\ha$ takes away an excitation: $\ha|n+1\rangle=\sqrt{n+1}|n\rangle$. 
A coherent state $|\alpha\rangle$, where $\alpha\in\mathbb{C}$,  
corresponds to an oscillation of the electromagnetic field, and it is the outcome of applying the displacement operator to the vacuum state, \ie $|\alpha\rangle=D(\alpha)|0\rangle$, 
which resembles the creation operation, with $D(\alpha)\equiv \exp(\alpha \ha^\dagger-\alpha^* \ha)$. 
 A thermal state $\tau(N)$ is a Gaussian mixture of coherent states,  where
\begin{align}
\tau(N)&\equiv \int_{\mathbb{C}} d^2 \alpha \frac{e^{-|\alpha|^2/N}}{\pi N} |\alpha\rangle \langle \alpha| 
\nonumber\\&
= \frac{1}{N+1}\sum_{n=0}^{\infty} \left(\frac{N}{N+1}\right)^n |n\rangle \langle n|
\label{eq:tau}
\end{align}
given an average photon number $N\geq 0$. 

Consider a bosonic MAC with cribbing encoders, whereby the channel input is a pair of electromagnetic field modes, with  annihilation operator $\ha_1$ and $\ha_2$, and the output is a modes with  annihilation operator $\hb$.
The annihilation operators  %
correspond to Alice 1, Alice 2, and Bob, respectively. %
The input-output relation of the bosonic MAC in the Heisenberg picture \cite{HolevoWerner:01p} is given by %
\begin{align}
\he &=\sqrt{\eta_1}\, \ha_1 +\sqrt{1-\eta_1}\,\hc \\
\ha_1'&=\sqrt{1-\eta_1}\, \ha_1 -\sqrt{\eta_1}\,\hc
\intertext{and}
\hb &=\sqrt{\eta_2}\, \ha_1' +\sqrt{1-\eta_2}\,\ha_2
\end{align}
where $\hc$ is associated with the environment noise and 
the parameter $\eta_k$  is  the transmissivity, $0\leq \eta_k\leq 1$, which captures, for instance, the length of the optical fiber and its absorption length \cite{EisertWolf:05c}. The relations above correspond to the outputs of a beam splitter, as illustrated in Figure~\ref{fig:BSp}. It is assumed that the encoder uses a coherent state protocol with an input constraint. That is, the input state  is a coherent state $|x_k\rangle$, $x_k\in\mathbb{C}$, for $k=1,2$, such that each codeword satisfies %
$\frac{1}{n}\sum_{i=1}^n |x_{k,i}|^2\leq N_{A_k}$.

Based on part 1 of Theorem~\ref{theo:cribSC}, we derive the following achievable region with strictly-causal cribbing, %
\begin{align}
&\mathcal{C}_{\text{s-c}}(\channel\circ\Lset)\supseteq %
\left\{ \begin{array}{rrl}
  (R_1,R_2) \,:\;
	&R_1 &\leq g(\eta_1 N_{A_1}+ (1-\eta_1) N_C) -g((1-\eta_1) N_C)  \\
  &R_2   &\leq g(\eta_2\eta_1 N_C+(1-\eta_2)N_{A_2})-g(\eta_2\eta_1 N_C)\\
	&R_1+R_2 &\leq g(  \eta_2(1-\eta_1) N_{A_1} +   \eta_2\eta_1 N_C+(1-\eta_2)N_{A_2})-g(\eta_2\eta_1 N_C)
	\end{array}
\right\}
\label{eq:inRscG}
\intertext{where $g(N)$ is the entropy of a thermal state with mean photon number $N$, }
g(x)&=\begin{cases}
(x+1)\log(x+1)-x\log(x) & x>0\\
0                       & x=0 \,.
\end{cases}
\end{align}
To obtain the region above, set the inputs to be mixed coherent states, $|X_1\rangle$ and $|X_2\rangle$, where $X_k$ is a circularly-symmetric Gaussian random variable with zero mean and variance $N_{A_k}/2$, for $k=1,2$, and let $U=0$. 

\begin{center}
\begin{figure}[tb]
\center
\includegraphics[scale=0.09,trim={2.5cm 0 25cm 2cm},clip]
{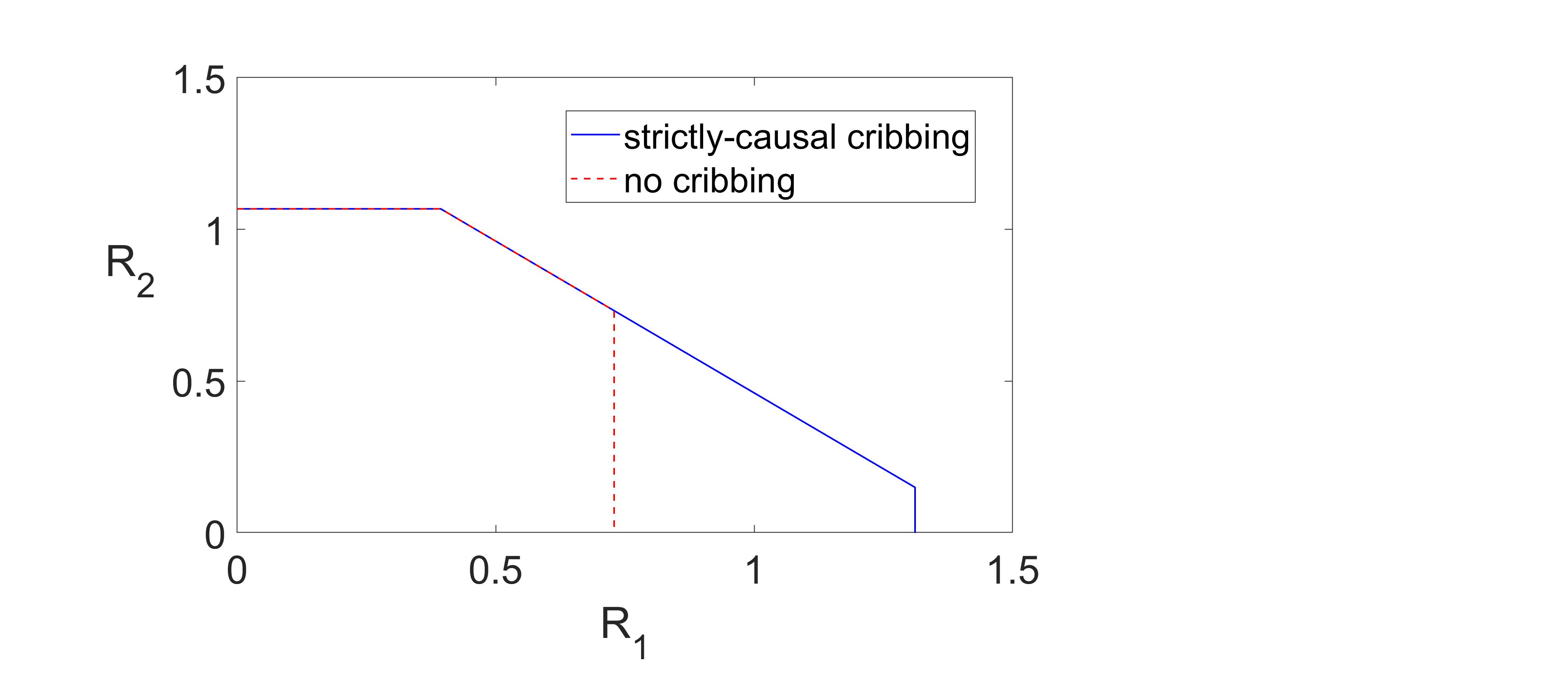} %
\caption{Achievable regions for the single-mode bosonic MAC. The decode forward achievable region with strictly-causal cribbing is the area below the thick blue line (see (\ref{eq:inRscG})). For comparison, the capacity region without cribbing $\mathcal{C}_{\text{none}}(\channel\circ\Lset)$ is depicted as the area below the red dashed line (see (\ref{eq:inR0G})). The transmission rate of Alice 1 can be significantly higher using cribbing.}
\label{fig:bosonicDF}
\end{figure}
\end{center}

On the other hand, the capacity region of single-mode Bosonic MAC without cribbing is \cite{YenShapiro:05p} 
\begin{align}
&\mathcal{C}_{\text{none}}(\channel\circ\Lset)=%
\left\{ \begin{array}{rrl}
  (R_1,R_2) \,:\;
	&R_1& \leq g(\eta_2(1-\eta_1)N_{A_1}+\eta_2\eta_1 N_C)-g(\eta_2\eta_1 N_C)  \\
  &R_2&   \leq g(\eta_2\eta_1 N_C+(1-\eta_2)N_{A_2})-g(\eta_2\eta_1 N_C)\\
	&R_1+R_2& \leq g(  \eta_2(1-\eta_1) N_{A_1} +   \eta_2\eta_1 N_C+(1-\eta_2)N_{A_2})-g(\eta_2\eta_1 N_C)
	\end{array}
\right\} \,.
\label{eq:inR0G}
\end{align}
The decode-forward achievable region with strictly-causal cribbing and the capacity region without cribbing are depicted in Figure~\ref{fig:bosonicDF} as the area below the thick blue line and below the dashed red line, respectively. As can be seen in the figure, cribbing can lead to a significant rate gain for Alice 1.

\vspace{-0.75cm}

\section{Partial Decode-Forward}
\label{sec:PDF}
We have pointed out in Remark~\ref{rem:robust} that if the cribbing system is not robust, then the decode-forward strategy in the previous section is not necessarily optimal. 
In fact, if the cribbing system $E$ that Alice 2 measures is noisier than the channel input $A_1'$, then the inner bounds above may be worse than in transmission without cribbing. This is easy to see when $E$ does not contain any information, say $|\Hset_E|=1$, in which case the decode-forward region leaves Alice 1 with zero rate,
\begin{align}%
\mathcal{R}^{\text{DF}}_{\text{s-c}}(\channel\circ\Lset)=\mathcal{R}^{\text{DF}}_{\text{caus}}(\channel\circ\Lset)=
\Big\{ (0,R_2) \,:\; R_2\leq \max_{p_{X}\,,\; \zeta_{A_2}^x} I(X;B)  \Big\} \,.
\end{align}%
In order to treat the case where the cribbing system is noisy,
we introduce a quantum %
cribbing method that is based on relay coding techniques (see Remark~\ref{rem:relay1} on the analogy between the relay channel and the cribbing model).
Specifically, we improve the inner bound by incorporating the partial decode-forward technique within our previous cribbing scheme.
Consider the quantum MAC  with strictly-causal cribbing.
 Define
\begin{align}%
\mathcal{R}^{\text{PDF}}_{\text{s-c}}(\channel\circ \Lset)=
\bigcup_{ p_{UV} p_{X_1 |U,V} p_{X_2 |U,V} \,,\; \theta_{A_1}^{x_1}\otimes \zeta_{A_2}^{x_2} }
\left\{ \begin{array}{rl}
  (R_1,R_2) \,:\;
	R_1 	&\leq I(V;E| %
	U)_\omega+ I(X_1;B|X_2 UV)_\omega  \\
  R_2   &\leq I(X_2;B|X_1 U)_\omega\\
	R_1+R_2 &\leq I(X_1 X_2;B)_\omega
	\end{array}
\right\}
\label{eq:inRscP}
\end{align}%
with
\begin{align}
&\omega_{UV X_1 X_2 A_1' E A_2}= \sum_{u,v,x_1,x_2} p_{U,V}(u,v)p_{X_1|U,V}(x_1|u,v)%
 p_{X_2|U,V}(x_2|u,v) \kb{u,v,x_1,x_2}%
\otimes \Lset_{A_1 \to A_1' E}(\theta_{A_1}^{x_1})\otimes \zeta_{A_2}^{x_2} \,,
\\
&\omega_{UV X_1 X_2 B}= \channel_{A_1' A_2\to B}(\omega_{UV X_1 X_2 A_1' A_2}) \,.
\end{align}
The superscript `PDF' stands for partial decode-forward coding.

\begin{theorem}
\label{theo:cribSCp}
Consider a quantum MAC $\channel_{A_1' A_2\to B}\circ \Lset_{A_1\to A_1' E}$. %
The rate region $\mathcal{R}^{\text{PDF}}_{\text{s-c}}(\channel\circ \Lset)$ is achievable for the quantum MAC  with strictly-causal cribbing, \ie
\begin{align}
 \mathcal{C}_{\text{s-c}}(\channel\circ\Lset)\supseteq \mathcal{R}^{\text{PDF}}_{\text{s-c}}(\channel\circ \Lset) \,.
\end{align}
\end{theorem}
The proof of Theorem~\ref{theo:cribSCp} is given in Appendix~\ref{app:cribSCp}. To the best of our knowledge, this is a new result for classical channels as well. As opposed to the analysis for Section~\ref{sec:main}, the decoder does not fully rely on the cribbing measurement to recover the information from Alice 1. Instead, we use rate-splitting such that part of Alice 1's message is decoded-forward through the cribbing system $E$, and the remaining part is decoded using $A_1'$.

For the classical-quantum MAC  with \emph{noisy} cribbing, we also prove a cutset upper bound.
In this case, the channel inputs are $A_1'\equiv A_1\equiv X_1$ and $A_2\equiv X_2$, and the cribbing channel $\Lset$ is represented by a classical noisy channel $Q_{Z|X_1}$ (see Definition~\ref{def:cqNoisy}).
 Then, we determine the capacity region in the special case where $Z$ is a determinisitic function of $X_1$, by showing that the partial decode-forward inner bound and the cutset outer bound coincide.
Given a classical-quantum MAC $\channel_{X_1 X_2\to B}\circ Q_{Z|X_1}$, define
\begin{align}%
\mathcal{R}^{\text{CS}}_{\text{s-c}}(\channel\circ Q)=%
\bigcup_{ p_{U} p_{X_1 |U} p_{X_2 |U} }
\left\{ \begin{array}{rl}
  (R_1,R_2) \,:\;
	R_1 	&\leq  I(X_1;BZ|X_2 U)_\omega  \\
  R_2   &\leq I(X_2;B|X_1 U)_\omega\\
	R_1+R_2 &\leq I(X_1 X_2;B)_\omega
	\end{array}
\right\}
\label{eq:inRscCS}
\end{align}
with
\begin{align}
&\omega_{U X_1 X_2 Z B}= \sum_{u,x_1,z,x_2} p_{U}(u)p_{X_1|U}(x_1|u) Q_{Z|X_1}(z|x_1) %
 p_{X_2|U}(x_2|u)  \kb{u,x_1,z,x_2}%
\otimes  \channel_{X_1 X_2\to B}(x_1,x_2) \,.
\end{align}
The superscript `CS' stands for the cutset outer bound.
\begin{theorem}
\label{theo:cribSCcs}
Consider a classical-quantum MAC %
with a noisy cribbing channel $Q_{Z|X_1}$. %
\begin{enumerate}[1)]
\item
The capacity region of the classical-quantum MAC $\channel_{X_1 X_2\to B}\circ Q_{Z|X_1}$   with strictly-causal noisy cribbing is bounded by
\begin{align}
 \mathcal{C}_{\text{s-c}}(\channel\circ Q)\subseteq \mathcal{R}^{\text{CS}}_{\text{s-c}}(\channel\circ Q) \,.
\end{align}
\item
If  $Q_{Z|X_1}$ is a $0$-$1$ matrix, \ie the cribbing observation $Z$ is a deterministic function of $X_1$, then
\begin{align}
 &\mathcal{C}_{\text{s-c}}(\channel\circ Q)=\mathcal{R}^{\text{PDF}}_{\text{s-c}}(\channel\circ Q)=\mathcal{R}^{\text{CS}}_{\text{s-c}}(\channel\circ Q)=
\bigcup_{ p_{U} p_{X_1 |U} p_{X_2 |U} }
\left\{ \begin{array}{l}
  (R_1,R_2) \,:\;\\
	R_1 	\leq H(Z|U)+ I(X_1;B|X_2 U Z)_\omega  \\
  R_2   \leq I(X_2;B|X_1 U)_\omega\\
	R_1+R_2 \leq I(X_1 X_2;B)_\omega
	\end{array}
\right\} \,.
\label{eq:inRscPcrib}
\end{align}
\end{enumerate}
\end{theorem}
The proof of Theorem~\ref{theo:cribSCcs} is given in Appendix~\ref{app:cribSCcs}.
Part 1 seems to be new for classical channels as well, while part 2 is the classical-quantum version of the classical result on the classical MAC with partial cribbing \cite{AsnaniPermuter:13p}.
Although,  Asnani and Permuter \cite{AsnaniPermuter:13p} considered a more complex network with cribbing at both encoders.

\section{Summary and Discussion}
\label{sec:summary}
We consider the quantum MAC $\Mset_{A_1 A_2\to B}$ with cribbing encoders. In quantum communication, the description is %
more delicate. 
By the no-cloning theorem, universal copying of quantum states is impossible.  Therefore, in the view of quantum mechanics, perfect  cribbing is against the laws of nature.
As illustrated in Figure~\ref{fig:Mcrib}, if Alice 1 sends $A_1$ through the channel, then Alice 2 is physically prohibited from having a copy of the input state. Hence, we consider the quantum MAC with noisy cribbing, consisting of a concatenation of a cribbing channel $\Lset_{A_1\to A_1' E}$ and the communication channel $\channel_{A_1' A_2\to B}$ (see Figure~\ref{fig:Mcrib}b).
Specifically, Alice 1 sends her input system $A_1$ through a cribbing channel that has two outputs, $A_1'$ and $E$.
Alice 2 performs a measurement on the system $E$, and uses the measurement outcome in order to encode the input state of $A_2$. Then, the input systems  $A_1'$ and $A_2$ are sent through the communication channel.
The model can also be interpreted as if the second transmitter performs a measurement on the environment $E$ of the first transmitter.

We consider the scenarios of strictly-causal, causal, and non-causal cribbing. In strictly-causal cribbing,  Alice 2 transmits $A_{2,i}$ at time $i$, and then measures the cribbing system $E_i$ after the transmission. Hence, she only knows the past measurement outcomes $z^{i-1}$ at time $i$. In the causal setting, Alice 2 measures the cribbing system $E_i$, at time $i$, before she transmits. Hence, she knows the past and present measurement outcomes $z^i=(z^{i-1},z_i)$ at time $i$. Whereas, in the non-causal case, Alice 2 can perform a joint measurement on $E^n$ a priori, \ie before the beginning of her transmission.
The entanglement between Transmitter 1 and the cribbing system of Transmitter 2 has the following implication.
In both the causal and non-causal scenarios, Alice 2 performs the measurement before 
the input $A_{1,i}'$ is sent through the channel. Hence, the cribbing measurement may inflict a ``state collapse" for Alice 1's transmission. In other words, in quantum communication, the cribbing operation interferes with Alice 1's input before it is even transmitted through the communication channel.
For a MAC with \emph{robust cribbing}, there exists a recovery channel from $E$ to $EA_1'$. Thereby,  the cribbing system $E$ includes all the information that is available in $A_1'$.
We derived achievable regions for each setting and established
a regularized capacity characterization  for robust cribbing.
%
%
%

%
%
%
%
%
%
Suppose Alice 1 and Alice 2 send three messages to Bob, a common message $m_0$ and two private messages $m_1$ and $m_2$. It is assumed that there is no cribbing.
The  quantum multiparty packing lemma \cite[Lemm. 3]{DingGharibyanHaydenWalter:20p} implies that the capacity region of the quantum MAC $\Mset_{A_1 A_2\to B}$ with a common message, without cribbing, is given by the regularization of the following region,
\begin{align}%
\inR_{\text{none}}^*(\Mset)=
\bigcup_{ p_U p_{X_1 |U} p_{X_2 |U} \,,\; \theta_{A_1}^{x_1}\otimes \zeta_{A_2}^{x_2} }
\left\{ \begin{array}{rl}
  (R_0,R_1,R_2) \,:\;
	R_1 &\leq I(X_1;B|X_2 U)_\omega  \\
  R_2   &\leq I(X_2;B|X_1 U)_\omega\\
	R_1+R_2 &\leq I(X_1 X_2;B|U)_\omega\\
	R_0+R_1+R_2 &\leq I(X_1 X_2;B)_\omega
	\end{array}
\right\}
\label{eq:inRnone0}
\end{align}%
where the union is as in (\ref{eq:inRnone}),  %
with %
$|\Uset|=|\Hset_{A_1}|^2+|\Hset_{A_2}|^2+1$, $|\Xset_1|\leq |\Uset|(|\Hset_{A_k}|^2+1)$, for $k=1,2$.
%
%
%
%
%
 %
%
%

The setting of noisy cribbing is significantly more challenging and it is closely-related to the relay channel. 
We further investigated the comparison between the relay channel and the MAC with cribbing encoders in Section~\ref{sec:PDF} (see Remark~\ref{rem:relay1}).  Building on the analogy between the noisy cribbing model and the relay channel, we developed a partial decode-forward region for the quantum MAC with strictly-causal non-robust cribbing.
For the classical-quantum MAC with cribbing encoders, we have completely determined the capacity region with perfect cribbing of the classical input, and derived a cutset region for noisy cribbing. In the special case of a classical-quantum MAC with a deterministic cribbing channel, the inner and outer bounds coincide.

Cooperation in terms of a quantum MAC offers additional advantages for resilience and trustworthiness, \eg to compensate for jamming attacks by an adversary \cite[Section VI.C.]{PeregSteinberg:21p}. As previously mentioned, in future communication systems, not only data but also physical and virtual objects will be controlled \cite{FettwisBoche:21m}. 
This growing demand of communication resources sharpens the need for quantum technology, and it is thus
 interesting to develop the corresponding theory.

\section*{Acknowledgment}
U. Pereg and H. Boche were supported 
by the German Research Foundation (DFG) under Germany's Excellence Strategy – EXC-2111 – 390814868. 
In addition,
U. Pereg, C. Deppe, and H. Boche were supported by
the German Federal
Ministry of Education and Research (BMBF) through Grants
16KISQ028 (Pereg, Deppe) and 
16KISQ020 (Boche). 
This work of H. Boche was supported in part by the BMBF within the national initiative for
``Post Shannon Communication (NewCom)" under Grant 16KIS1003K, and in
part by the DFG within the Gottfried Wilhelm
Leibniz Prize under Grant BO 1734/20-1 and within Germany's Excellence
Strategy EXC-2092 – 390781972. U. Pereg was also supported by the Israel CHE Fellowship for Quantum Science and Technology.

\begin{appendices}

\section{Proof of Lemma~\ref{lemm:CardCl} (Cardinality Bounds)}
\label{app:CardCl}
To bound the alphabet size of the random variables $U$, $X_1$, and $X_2$, we use the Fenchel-Eggleston-Carath\'eodory lemma \cite{Eggleston:66p} and similar arguments as in  \cite{YardHaydenDevetak:08p,
Pereg:22p}.
Let
\begin{align}
d_0&= |\Hset_{B}|^2+2\\
d_1&=|\Uset|(|\Hset_{A_1}|^2+2) \label{eq:L0} \\
d_2&=|\Uset|(|\Hset_{A_2}|^2+1) \,. \label{eq:L1}
\end{align}

First, we bound the cardinality of $\Uset$. Consider a given ensemble $\{ p_{X_1|U}(x_1|u)p_{X_2|U}(x_2|u) \,, \theta_{A_1}^{x_1}\otimes \zeta_{A_2}^{x_2} \}$.
Every density matrix $\rho_A$ has a unique parametric representation $\mathscr{V}(\rho_A)$ of dimension %
 $|\Hset_A|^2-1$. 
Then, define a map $f_0:\Uset\rightarrow \mathbb{R}^{d_0}$ by
\begin{align}%
f_{0}(u)= \Big(  \mathscr{V}(\omega_B^u) \,,\; H(B|X_1,X_2,U=u)_\omega 
 \,,\;  I(X_1;E| U=u)_\omega   \,,\; I(X_2;B|X_1 U=u)_\omega    \Big)
\end{align}%
where $\omega_{B}^{u}=\sum_{x_1,x_2}  p_{X_1|U}(x_1|u)p_{X_2|U}(x_2|u) $ $\channel\left( \trace_E\left(\Lset(\theta_{A_1}^{x_1})\right)\otimes \zeta_{A_2}^{x_2}\right) $. The map $f_0$ can be extended to a map that  acts on probability distributions as follows,
\begin{align}%
F_{0} \,:\; p_{U}(\cdot)  \mapsto
\sum_{u\in\Uset} p_{U}(u) f_{0}(u)= \Big(  \mathscr{V}(\omega_B) \,,\;  H(B|X_1 X_2)_\omega \,,\; I(X_1;E| U)_\omega   \,,\; I(X_2;B|X_1 U)_\omega   \Big)  \,.
\end{align}%
According to the Fenchel-Eggleston-Carath\'eodory lemma \cite{Eggleston:66p}, any point in the convex closure of a connected compact set within $\mathbb{R}^d$ belongs to the convex hull of $d$ points in the set. 
Since the map $F_{0}$ is linear, it maps  the set of distributions on $\Uset$ to a connected compact set in $\mathbb{R}^{d_0}$. %
Thus, for every  $p_{U}$, 
there exists a probability distribution $p_{\bar{U}}$ on a subset $\bar{\Uset}\subseteq \Uset$ of size $%
d_0$, such that 
$%
F_{0}(p_{{\bar{U}}})=F_{0}(p_{U}) %
$. %
We deduce that alphabet size can be restricted to $|\Uset|\leq d_0$, while preserving $\omega_{B}$, $H(B|X_1 X_2)_\omega$, $I(X_1;E|U)_\rho$, 
$I(X_2;B|X_1 U)_\rho$,  and thus, $H(B)_\omega$ and
$I(X_1 X_2;B)_\omega=H(B)_\omega-H(B|X_1 X_2)_\omega$. 

Next, we bound the alphabet size for the auxiliary variables $X_1$ and $X_2$.
For every $u\in\Uset$, define a map $f_{1,u}:\Xset_1\rightarrow \mathbb{R}^{|\Hset_{A_1}|^2+2}$ by
\begin{align}%
f_{1,u}(x_1)= \Big(  \mathscr{V}(\theta_{A_1}^{x_1}) \,,\; H(E|X_1=x_1, U=u)_\omega   \,,\; H(B|X_1=x_1, U=u)_\omega \,,\; H(B|X_2,X_1=x_1, U=u)_\omega    \Big) \,.
\end{align}%
Then, the map $f_1$ is extended to %
\begin{align}%
F_{1,u} \,:\; p_{X_1|U}(\cdot|u)  \mapsto &
\sum_{x_1\in\Xset_1} p_{X_1|U}(x_1|u) f_{1,u}(x_1)= \nonumber\\& \Big(  \mathscr{V}(\theta^u_{A_1}) \,,\;  H(E|X_1, U=u)   \,,\; H(B|X_1, U=u)_\omega \,,\; H(B|X_2,X_1, U=u)_\omega     \Big) %
\end{align}%
with
\begin{align}
\theta_{A_1}^u\equiv \sum_{x_1} p_{X_1|U}(x_1|u) \theta_{A_1}^{x_1} \,.
\label{eq:thetaU}
\end{align} 
Thus, by Fenchel-Eggleston-Carath\'eodory lemma \cite{Eggleston:66p}, for every  $p_{X_1|U}(\cdot|u)$, 
there exists a probability distribution $p_{\bar{X}_1|U}(\cdot|u)$ on a subset $\bar{\Xset}_1\subseteq \Xset_1$ of size $%
|\Hset_{A_1}|^2+2$, such that 
$%
F_{1,u}(p_{{\bar{X}_1|U}}(\cdot|u))=F_{1,u}(p_{X_1|U}(\cdot|u))) %
$. %
We deduce that alphabet size can be restricted to $|\Xset_1|\leq d_1$, while preserving  $H(E|X_1 U)$, $H(B|X_1 U)_\omega$, $H(B|X_1 X_2)_\omega$, and  
\begin{align}
&\omega_{U A_1 X_2 A_2}=\sum_u p_U(u)p_{X_2|U}(x_2|u) \kb{u}  %
\otimes \theta_{A_1}^u\otimes \kb{x_2}\otimes \zeta^{x_2}_{A_2} 
\end{align}
where $\theta_{A_1}^u$ is as in (\ref{eq:thetaU}).
 This implies that $\omega_{U A_1' E X_2 A_2}=\Lset(\omega_{U A_1 X_2 A_2})$,  $H(E|U)_\omega$, and $I(X_1;E|U)_\omega=H(E|U)_\omega-H(E|X_1 U)_\omega$ remain the same, and also
$\omega_{U X_2 B}\equiv \channel_{A_1' A_2\rightarrow B}(\omega_{U X_2 A_1' A_2})$, $H(B)_\omega$, $I(X_2;B|X_1 U)_\omega=H(B|X_1 U)_\omega-H(B|X_1 X_2)_\omega$,
 and $I(X_1,X_2;B)_\omega=H(B)_\omega-H(B|X_1,X_2)_\omega$.
The bound  $|\Xset_2|\leq d_2$ follows from the same argument using the function $f_{2,u}:\Xset_2\rightarrow \mathbb{R}^{|\Hset_{A_2}|^2+1}$,
\begin{align}%
f_{2,u}(x_2)= \Big(  \mathscr{V}(\zeta_{A_2}^{x_2}) \,,\; H(B|X_1,X_2=x_2, U=u)_\omega   \,,\; H(B|X_1,X_2=x_2)_\omega   \Big) 
\end{align}%
for $u\in\Uset$. This completes the proof for the cardinality bounds.
\qed

\section{Proof of Theorem~\ref{theo:cribSC}}
\label{app:cribSC}
Consider the quantum MAC $\channel\circ\Lset$ with strictly-causal cribbing.

\subsection*{Part 1}
We show that for every $\delta_1,\delta_2,\eps_0>0$, there exists a $(2^{n(R_1-\delta_1)},2^{n(R_2-\delta_2)},n,\eps_0)$ code for $\channel_{A_1' A_2\rightarrow B}\circ \Lset_{A_1\to A_1' E}$ with strictly-causal cribbing at Encoder 2, provided that $(R_1,R_2)\in \mathcal{R}^{\text{DF}}_{\text{s-c}}(\channel\circ\Lset)$. 
To prove achievability, we extend  the classical block Markov coding with backward decoding to the quantum setting, and then apply the quantum packing lemma.

We use $T$ transmission blocks, where each block consists of $n$ channel uses. In particular, with strictly-causal cribbing, Encoder 2 has access to the cribbing measurements from the previous blocks. Each transmitter sends $T-1$ messages. Let us fix $m_k(0)=m_k(T) \equiv 1$ for $k=1,2$. Alice 1 sends $(m_1(j))_{j=1}^{T-1}$, and Alice 2 $(m_2(j))_{j=1}^{T-1}$.
 Hence, the coding rate for User $k$ is
$\left(\frac{T-1}{T}\right) R_k$, which tends to $R_k$ as the number of blocks $T$ grows to infinity.

Let $\{ p_U(u)p_{X_1|U}(x_1|u) p_{X_2|U}(x_2|u), \theta^{x_1}$ $\otimes \zeta^{x_2} \}$ be a  given ensemble over $\Hset_{A_1}\otimes \Hset_{A_2}$. 
Define 
\begin{align}
\omega_{A_1'E}^{x_1}&= \Lset_{A_1\to A_1'E}( \theta^{x_1}) \,,\\
\omega_B^{x_1,x_2}&= \channel_{A_1' A_2\to B} (\omega_{A_1'}^{x_1}\otimes \zeta^{x_2})%
\,.
\end{align}
 In addition, let
\begin{align}
\theta^u&=\sum_{x_1\in\Xset_1} p_{X_1|U}(x_1|u) \theta^{x_1}
\\
\zeta^u &=\sum_{x_2\in\Xset_2} p_{X_2|U}(x_2|u) \zeta^{x_2}
\\
\omega_{A_1' E}^{u}&=\Lset_{A_1\to A_1'E}( \theta^{u})
\\
\omega_B^{u}&= \channel_{A_1' A_2\to B} (\omega_{A_1'}^{u}\otimes \zeta^{u})
\end{align}
for $u\in\Uset$.
Hence, $\omega_{A_1' E}$ and $\omega_B$ are the corresponding average states.

\vspace{0.2cm}
The code construction, encoding with cribbing, and decoding procedures are described below.
\vspace{0.2cm}
\subsection{Classical Codebook Construction}
$\,$
\begin{enumerate}[(i)]
\item
Generate $2^{nR_1}$ independent sequences $u^n(m_0)$, $m_0\in [1:2^{nR_1}]$, at random 
according to $\prod_{i=1}^n p_U(u_{i})$.
\item
For every $m_0$, generate $2^{nR_1}$ conditionally independent sequences $x_1^n(m_0,m_1)$, $m_1\in [1:2^{nR_1}]$, %
according to $\prod_{i=1}^n p_{X_1|U}(x_{1,i}|u_{i}(m_0))$.
\item
For every $m_0$, generate $2^{nR_2}$ conditionally independent sequences $x_2^n(m_0,m_2)$, $m_2\in [1:2^{nR_2}]$, %
according to $\prod_{i=1}^n p_{X_2|U}(x_{2,i}|u_{i}(m_0))$.
\end{enumerate}

\subsection{Encoding and Decoding}

\subsubsection*{Encoder 1}
To send the messages $(m_1(j))_{j=1}^{T-1}$,  Alice 1 performs the following.
In block $j$, set  
\begin{align}
&m_0=m_1(j-1) \,,\; m_1=m_1(j) \,.
\intertext{Then, prepare the state}
&\rho_{A_1^n(j)}=\theta^{x_1^n(m_0,m_1)}
\end{align}
 and send $A_1^n(j)$,  for $j\in [1:T]$.
As the $j$th transmission goes through the cribbing channel $\Lset_{A_1\to A_1' E}$, we have 
\begin{align}
\rho_{A_1'^n(j) E^n(j)}=\omega_{A_1'^n E^n}^{x_1^n(m_0,m_1)} \,.%
\end{align}
The second transmitter can access the cribbing system $E^n(j)$, which are entangled with Alice 1's transmission. 
Cribbing is performed by a sequence of measurements that recover the messages of Alice 1.
In each block, the choice of the cribbing measurement depends on the outcome in the previous block.
With strictly-causal encoding, Alice 2 must perform the cribbing measurement at the end of the block, after she has already sent $A_2^n(j)$.

\subsubsection*{Encoder 2}
To send the messages $(m_2(j))_{j=1}^{T-1}$,  Alice 2 performs the following. 
Fix $\tm_1(0)\equiv 1$. In block $j$, 
 given the previous cribbing estimate $\tm_1(j-1)$,  do  as follows, for $j=1,2,\ldots, T$.

\begin{enumerate}[(i)]
\item
Set
\begin{align}
&m_0=\tm_1(j-1) \,,\; m_2=m_2(j) \,.
\end{align}

\item 
Prepare the state
\begin{align}
&\rho_{A_2^n(j)}=\zeta^{x_2^n(m_0,m_2)}
\end{align}
 and send $A_2^n(j)$, for $j\in [1:T]$.

\item
 Measure the next cribbing estimate $\tm_1(j)$ by applying a POVM $\Kset_{E^n(j)|m_0}=\{ K_{m_1|m_0} \}$ that will be chosen later. %

\end{enumerate}

\subsubsection*{Backward decoding}
	The decoder recovers the messages using sequential measurements as well. Yet, the order is backwards, \ie the measurement of the $j$th message of Alice 1 is chosen based on the estimate of $m_1(j+1)$. Fix $\hm_1(0)=\hm_1(T)\equiv 1$.
In block $j$,  for $j=T,T-1,\ldots,2$, the decoder uses the previous estimate of $\hm_1(j)$, and 
  measures $\hm_1(j-1)$ and $\hm_2(j)$ using a POVM 
	$\Dset_{B^n(j)|\hm_1}=\{  D%
	_{m_0,m_2|\hm_1} \}$ with $\hm_1=\hm_1(j)$. Finally, in block $1$, the decoder measures $\hm_2(1)$ with a POVM 
	$\Dset_{B^n(1)|m_0,m_1}=\{  D'%
	_{m_2|\hm_0,\hm_1} \}$, with $\hm_0=1$ and $\hm_1=\hm_1(1)$. The decoding POVMs will also be specified later.

\subsection{Analysis of Probability of Error}
Let $\delta>0$. We use the notation $\eps_i(\delta)$, $i=1,2,\ldots$,
for terms that tend to zero as $\delta\rightarrow 0$.
By symmetry, we may assume without loss of generality that the transmitters send the messages $M_1(j)=M_2(j)=1$.
Consider the following events,
\begin{align}
\mathscr{E}_0(j)=& \{  (U^n(M_1(j-1)),X_1^n(M_1(j-1),1),%
X_2^n(M_1(j-1),1))\notin \Aset^{\delta_1}(p_{U,X_1,X_2}) \} \\
\mathscr{E}_1(j)=& \{  \tM_1(j)\neq 1  \}\\
\mathscr{E}_2(j)=& \{  \hM_1(j-1) \neq 1  %
\}
\\
\mathscr{E}_3(j)=& \{  \hM_2(j)\neq 1 \} 
\end{align}
for $j\in [1:T]$, with $\delta_1\equiv \delta/(2 |\Xset_2| |\Uset|)$.
By the union of events bound, the probability of error is bounded by
\begin{align}
P_{e|m_1(j)=m_2(j)=1}^{(Tn)}(\Fset_1,\Fset_2,\Kset,\Dset) %
&\leq %
\sum_{j=1}^{T} \prob{ \mathscr{E}_0(j) }%
+\sum_{j=1}^{T} \cprob{ \mathscr{E}_1(j) }{ \mathscr{E}_0^c(j)\cap \mathscr{E}_1^c(j-1) } \nonumber\\&
+\sum_{j=1}^{T}\cprob{ \mathscr{E}_2(j)\cup \mathscr{E}_3(j) }{ \mathscr{E}_0^c(j)\cap \mathscr{E}_1^c(j) \cap \mathscr{E}_2^c(j+1) } %
\label{eq:PeBsc}
\end{align}
where the conditioning on $M_1(j)=M_2(j)=1$ is omitted for convenience of notation.
By the weak law of large numbers, the probability terms $\prob{ \mathscr{E}_0(j) }$ tend to zero as $n\rightarrow\infty$.

To bound the second sum, which is associated with the cribbing measurements, we use the quantum packing lemma.
Alice 2's measurement is effectively a decoder for the marginal cribbing channel $\Lset^{(1)}_{A_1\to E}$, which is defined by 
$\Lset^{(1)}_{A_1\to E}(\rho_{A_1})\equiv \trace_{A_1'}\left( \Lset_{A_1\to E A_1'}(\rho_{A_1}) \right) $.
 Given $\mathscr{E}_0^c(j)$, we have that $(U^n(M_1(j-1)),X_1^n(M_1(j-1),1))\in \Aset^{\delta}(p_{U,X_1}) $. %
Now, observe that
\begin{align}
\trace\left[ \Pi^{\delta}(\omega_E|u^n,x_1^n) \omega_{E^n}^{u^n,x_1^n} \right] \geq& 1-\eps_1(\delta) 
\\
\Pi^{\delta}(\omega_E|u^n)  \omega_{E^n}^{u^n}   \Pi^{\delta}(\omega_E|u^n) \preceq& 2^{ -n(H(E|U)_{\omega}-\eps_1(\delta)) } \Pi^{\delta}(\omega_E)
\\
\trace\left[ \Pi^{\delta}(\omega_E|u^n,x_1^n)  \right] \leq& 2^{ n(H(E|U,X_1)_{\omega} +\eps_1(\delta))} \\
\trace\left[ \Pi^{\delta}(\omega_E|u^n) \omega_{E^n}^{u^n,x_1^n} \right] \geq& 1-\eps_1(\delta) 
\end{align}
for all $(u^n,x_1^n)\in\Aset^{\delta}(p_{U,X_1})$, by 
(\ref{eq:UnitTCond})-(\ref{eq:UnitTCondB}), respectively.
Since the codebooks are statistically independent of each other, we have by the single-user quantum packing lemma \cite{HsiehDevetakWinter:08p}   \cite[Lemma 12]{Pereg:22p}, that there exists a POVM $K_{m_1|u^n}$ such that
$%
\cprob{ \mathscr{E}_1(j) }{ \mathscr{E}_0^c(j)\cap \mathscr{E}_1^c(j-1) } \leq 2^{ -n( I(X_1;E|U)_\rho -R_1-\eps_2(\delta)) } 
$, %
which tends to zero as $n\rightarrow\infty$, provided that 
\begin{align}
R_1< I(X_1;E|U)_\omega -\eps_2(\delta) \,.
\label{eq:Er1}
\end{align}

We move to the last sum in the RHS of (\ref{eq:PeBsc}). %
Here, we use the quantum multi-party packing lemma \cite{DingGharibyanHaydenWalter:20p}. 
 Suppose that $\mathscr{E}_1^c(j)\cap \mathscr{E}_2^c(j+1)$ occurred, namely Encoder 2 measured the correct $M_1(j-1)$ and the decoder measured the correct $M_1(j)$. 
Then, we set $R_0\leftarrow R_1$ and $R_1\leftarrow 0$ in the quantum multi-party packing lemma, since we are decoding $(m_0,m_2)\equiv (M_1(j-1), M_2(j))$ while conditioning on $m_1\equiv M_1(j)$. 
By %
\cite[Lemm. 3]{DingGharibyanHaydenWalter:20p}, 
there exists a POVM $D_{m_0,m_2|m_1}$ such that the probability term $\cprob{ \mathscr{E}_2(j)\cup \mathscr{E}_3(j) }{ \mathscr{E}_0^c(j)\cap \mathscr{E}_1^c(j) \cap \mathscr{E}_2^c(j+1) }$ tends to zero as $n\to\infty$
provided that 
\begin{align}
R_2&< I(X_2;B|X_1 U)_\omega-\eps_4(\delta)
\label{eq:B30}
\intertext{and}
R_1+R_2&<I(X_1 X_2;B)_\omega-\eps_4(\delta) \,.
\label{eq:B31}
\end{align}
This completes the achievability proof.

\subsection*{Part 2}
Consider the quantum MAC with strictly-causal \emph{robust}  cribbing. 
To show that rate pairs in $\frac{1}{\kappa}\mathcal{R}^{\text{DF}}_{\text{s-c}}((\channel\circ\Lset)^{\otimes \kappa})$ are achievable, one may employ the coding scheme from part 1 for the product MAC $(\channel\circ\Lset)^{\otimes \kappa}$, where $\kappa$ is arbitrarily large.
Now, we show the converse part using standard considerations along with the  quantum Markov chain property for robust cribbing (see Definition~\ref{def:robust}). %

 Suppose that Alice 1 chooses $m_1$ uniformly at random, and prepares an input state $\rho^{m_1}_{ A_1^n}$. 
Upon sending the systems $A_1^n$ through the cribbing channel,  we have
$\rho^{m_1}_{A_1'^n E^n}=  \Lset_{A_1^n\rightarrow A_1'^n E^n}(\rho^{m_1}_{ A_1^n}) $.
Before preparing the state of her system $A_{2,i}$, Alice 2 can measure the cribbing systems $ {E^{i-1}} $, and obtain  an outcome $z_{i-1}$.
Hence, Alice 2 prepares the input state $\rho^{m_2,z^{i-1}}_{ A_2^i}$. Then, $A_{1,i}'$ and $A_{2,i}$ are sent through the MAC $\channel_{A_1' A_2\to B}$. Bob receives the output systems $B^n$ and performs a measurement in order to obtain an estimate $(\hm_1,\hm_2)$ of the message pair.

Consider a sequence of codes $(\Fset_{1n},\Fset_{2n},\Kset_n,\Dset_n)$ such that the average probability of error tends to zero, hence
the error probabilities $\prob{ \hM_1\neq M_1 }$, $\prob{ (\hM_1,\hM_2)\neq (M_1,M_2)}$, $\prob{ \hM_2\neq M_2 |M_1}$ are bounded by some
$\alpha_n$ which tends to zero as $n\rightarrow \infty$.
By Fano's inequality \cite{CoverThomas:06b}, it follows that%
\begin{align}
H(M_1|\hM_1) \leq n\eps_n\\
H(M_1,M_2|\hM_1,\hM_2) \leq n\eps_n'\\
H(M_2|\hM_2,M_1) \leq n\eps_n''
\label{eq:AFWC2c}
\end{align}
where $\eps_n,\eps_n',\eps_n''$ tend to zero as $n\rightarrow\infty$.
Hence, 
\begin{align}
nR_1&= H(M_1)=I(M_1;\hM_1)_{\rho}+H(M_1|\hM_1) 
\nonumber\\
&\leq I(M_1;\hM_1)_{\rho}+n\eps_n \nonumber\\
&\leq I(M_1;B^n )_{\rho}+n\eps_n \nonumber\\
&\leq I(M_1;A_1'^n E^n)_{\rho}+n\eps_n 
\label{eq:ConvIneq1SC}
\end{align}
where the second inequality %
follows from the Holevo bound (see Ref. \cite[Theo. 12.1]{NielsenChuang:02b}), and the last inequality follows from the data processing inequality. Now,  given robust cribbing, the systems $M_1\Cbar E^n\Cbar A_1'^n$ form a quantum Markov chain, by  Definition~\ref{def:robust}.
 Hence, 
$I(M_1;A_1'^n| E^n)_{\rho}=0$ and $I(M_1;A_1'^n E^n)_{\rho}=I(M_1;E^n)_{\rho}$. Thus,
\begin{align}
nR_1&\leq I(M_1;E^n)_{\rho}+n\eps_n .
\label{eq:ConvIneq1SCb}
\end{align}

By the same considerations, we also have
\begin{align}
n(R_1+R_2)&= %
I(M_1 M_2;\hM_1,\hM_2)_{\rho}+H(M_1 M_2|\hM_1,\hM_2) 
\nonumber\\
&\leq I(M_1 M_2;B^n)_{\rho}+n\eps_n '
\label{eq:ConvIneq3SC}
\intertext{and}
nR_2&= %
I(M_2;\hM_2|M_1)_{\rho}+H(M_2|\hM_2,M_1) 
\nonumber\\
&\leq I(M_2;B^n|M_1 )_{\rho}+n\eps_n '' \,.
\label{eq:ConvIneq2SC}
\end{align}
The proof for the regularized region follows from (\ref{eq:ConvIneq1SCb})-(\ref{eq:ConvIneq2SC}) by defining $X_1^n=f_1(M_1)$, $X_2^n=f_2(M_2)$, and $U^n=\emptyset$, where $f_k$ is an arbitrary one-to-one map from $[1:2^{nR_k}]$ to $\Xset_k^n$, for $k=1,2$. 
\qed

\section{Proof of Corollary~\ref{coro:cribSCcq}}
\label{app:cribSCcq}
Consider the classical-quantum MAC $\channel_{X_1 X_2\to B}\circ \mathrm{id}_{X_1\to X_1 X_1}$ with strictly-causal noiseless cribbing. 
Since the cribbing system stores a perfect copy of Alice 1's classical input in this case, \ie $E\equiv X_1$, we have $I(X_1;E|U)_\omega=H(X_1|U)$. Thereby, achievability immediately follows from part 1 of Theorem~\ref{theo:cribSC}.
Note that the classical-quantum setting with noiseless cribbing is a special case of robust cribbing. Hence, we can use the derivation of part 2 of Theorem~\ref{theo:cribSC} as well.
 
As for the converse proof, let $X_1^n\equiv \Fset_1(M_1)$ and $X_{2,i}\equiv \Fset_{2,i}(M_2,X_1^{i-1})$ denote the channel inputs. 
Since the encoding map $\Fset_1$ is  classical, there exists a random element
$S_1$ that controls the encoding function. That is, $X_1^n\equiv f_1(M_1,S_1)$, where 
$f_1$ is a deterministic function, and $S_1$ is statistically independent of the message. Define
\begin{align}
U_i\equiv (X_1^{i-1},S_1) \,.
\label{eq:convSCcqUi}
\end{align}
Then, the transmission rate of Alice 1 satisfies
\begin{align}
nR_1&= H(M_1)
\nonumber\\
&= H(M_1|S_1)
\nonumber\\
&= H(X_1^n|S_1)
\nonumber\\
&= \sum_{i=1}^n H(X_{1,i}|U_i)
\label{eq:ConvIneq1SCbcq}
\end{align}
where the second equality holds since $S_1$ is independent of the message,
the third since $X_1^n$ is a deterministic function of $(M_1,S_1)$, and the last equality follows from the entropy chain rule (see (\ref{eq:convSCcqUi})).

Next, we bound the transmission rate of Alice 2 as follows,
\begin{align}
nR_2&=H(M_2) \nonumber\\
&= H(M_2|M_1, S_1)
\nonumber\\
&= I(M_2;\hM_2|M_1, S_1)+H(M_2|\hM_2, M_1, S_1) 
\nonumber\\
&\leq I(M_2;\hM_2|M_1,S_1)+H(M_2|\hM_2,M_1) 
\nonumber\\
&\leq I(M_2;B^n|M_1,S_1 )_\rho+n\eps_n '' 
\label{eq:ConvIneq2SCbcq1}
\end{align}
where the first inequality holds since conditioning cannot increase entropy, and the second follows from the data processing inequality and Fano's inequality (see (\ref{eq:AFWC2c})). Using the chain rule, the last bound can be expressed as
\begin{align}
nR_2&\leq \sum_{i=1}^n I(M_2;B_i|M_1,S_1,B^{i-1} )_\rho+n\eps_n '' 
\nonumber\\
&= \sum_{i=1}^n I(M_2;B_i|M_1,S_1, X_{1,i}, X_1^{i-1}, B^{i-1} )_\rho+n\eps_n ''
\label{eq:ConvIneq2SCbcq2} 
\end{align}
 as $X_1^n$ is a deterministic function of $(M_1,S_1)$. Observe that the mutual information summand is then bounded by
\begin{align}
 &I(M_2;B_i|M_1,S_1, X_{1,i}, X_1^{i-1}, B^{i-1} )_\rho
\leq I(M_2,X_{2,i};B_i|M_1,S_1, X_{1,i}, X_1^{i-1}, B^{i-1} )_\rho
\nonumber\\
&= H(B_i|M_1,S_1, X_{1,i}, X_1^{i-1}, B^{i-1} )_\rho %
- H(B_i|X_{1,i}, X_{2,i}, M_1,M_2,S_1,  X_1^{i-1}, B^{i-1} )_\rho
\nonumber\\
&\leq H(B_i|S_1, X_{1,i}, X_1^{i-1} )_\rho %
- H(B_i|X_{1,i}, X_{2,i}, M_1,M_2,S_1,  X_1^{i-1}, B^{i-1} )_\rho \,.
\label{eq:ConvIneq2SCbcq3}
\end{align}
Consider the second term, and observe that given 
$X_{1,i}=x_{1,i}$, $X_{2,i}=x_{2,i}$, the output system $B_i$ is in the state
$\channel(x_{1,i},x_{2,i})$ and it has no correlation with $M_1$, $M_2$, $S_1$,  $X_1^{i-1}$, and $B^{i-1}$. That is, given $X_{1,i}$ and $X_{2,i}$, the 
system $B_i$ is in a product state with the joint system of $(M_1,M_2,S_1,X_1^{i-1},B^{i-1})$.
Thus, the second term equals $H(B_i| X_{1,i}, X_{2,i} )_\rho$. Similarly, $H(B_i|S_1, X_{1,i}, X_{2,i}, X_1^{i-1} )_\rho=H(B_i| X_{1,i}, X_{2,i} )_\rho$ as well, which implies
\begin{align}
&H(B_i|X_{1,i}, X_{2,i}, M_1,M_2,S_1,  X_1^{i-1}, B^{i-1} )_\rho %
=H(B_i|X_{1,i}, X_{2,i}, S_1,  X_1^{i-1} )_\rho \,.
\label{eq:ConvIneq2SCbcq4}
\end{align}
Therefore, by  (\ref{eq:ConvIneq2SCbcq2})-(\ref{eq:ConvIneq2SCbcq4}), along with the definition of $U_i$ in (\ref{eq:convSCcqUi}),
\begin{align}
nR_2&\leq \sum_{i=1}^n \left[ H(B_i| X_{1,i}, U_i )_\rho-H(B_i| X_{1,i}, X_{2,i}, U_i )_\rho  \right]+n\eps_n ''
\nonumber\\
&= \sum_{i=1}^n I(X_{2,i};B_i| X_{1,i}, U_i )_\rho+n\eps_n '' \,.
\label{eq:ConvIneq2SCbcq5}
\end{align}

The bound on the sum-rate is straightforward. Indeed, by (\ref{eq:ConvIneq3SC}),
\begin{align}
n(R_1+R_2)
&\leq 
\sum_{i=1}^n I(M_1 M_2;B_i|B^{i-1})_{\rho}+n\eps_n '
\nonumber\\
&\leq \sum_{i=1}^n I(M_1 M_2 B^{i-1} ;B_i)_{\rho}+n\eps_n '
\nonumber\\
&\leq \sum_{i=1}^n I(M_1 M_2 B^{i-1} X_{1,i} X_{2,i} ;B_i)_{\rho}+n\eps_n '
\nonumber\\
&= \sum_{i=1}^n I(X_{1,i} X_{2,i} ;B_i)_{\rho}+n\eps_n '
\label{eq:ConvIneq3SCbcq}
\end{align}
since $I(M_1 M_2 B^{i-1} ;B_i|X_{1,i} X_{2,i} )_{\rho}=0$.

To complete the proof, consider a time index $K$ that is drawn uniformly at random, from $[1:n]$, independently of $M_1$, $M_2$, and $S_1$. Then, by (\ref{eq:ConvIneq1SCbcq}), (\ref{eq:ConvIneq2SCbcq5}), and (\ref{eq:ConvIneq3SCbcq}),
\begin{align}
R_1&=H(X_{1,K}|U_K,K)
\label{eq:ConvIneq1SCbcq2}
\\
R_2-\eps_n '' &\leq I(X_{2,K};B_K| X_{1,K}, U_K,K )_\rho
\label{eq:ConvIneq2SCbcq6}
\\
R_1+R_2-\eps_n ' &\leq I(X_{1,K} X_{2,K} ;B_K|K)_{\rho}  \,.
\label{eq:ConvIneq3SCbcq2}
\end{align}
Furthermore, define a joint state $\omega_{UX_1 X_2 B}$ by identifying $U\equiv (K,U_K)$, $X_1\equiv X_{1,K}$, $X_2\equiv X_{2,K}$, and
$B\equiv B_K$. That is,
\begin{align}
&\omega_{UX_1 X_2 B}\equiv \sum_{i=1}^n \sum_{u_i,x_{1},x_{2}} \frac{1}{n} p_{U_i}(u_i)p_{X_{1,i}|U_i}(x_{1}|u_i)%
\cdot p_{X_{2,i}|U_i}(x_{2}|u_i) \kb{i,u_i} \otimes \kb{x_{1}}\otimes \kb{x_{2}}%
\otimes \channel(x_{1},x_{2}) \,.
\end{align}
Thus, the individual rates are bounded by
\begin{align}
R_1&= H(X_1|U)
\\
R_2-\eps_n '' &\leq I(X_{2};B| X_{1}, U )_\omega \,.
\end{align}
As for the sum-rate bound in (\ref{eq:ConvIneq3SCbcq2}),
\begin{align}
R_1+R_2-\eps_n ' &\leq I(X_{1} X_{2} ;B|K)_{\omega} 
\nonumber\\
&\leq I(X_{1} X_{2} K ;B)_{\omega} 
\nonumber\\
&= I(X_{1} X_{2} ;B)_{\omega} +I(K;B|X_{1} X_{2})_\omega
\nonumber\\
&= I(X_{1} X_{2} ;B)_{\omega} 
\end{align}
as $I(K;B|X_{1} X_{2})_\omega=0$ since the channel has a memoryless product form.
This completes the proof of Corollary~\ref{coro:cribSCcq}.
\qed

\section{Proof of Theorem~\ref{theo:cribC}}
\label{app:cribC}
\subsection*{Part 1}
Consider the quantum MAC $\channel\circ\Lset$ with causal cribbing. 
Since the achievability proof is similar to the derivation for the strictly-causal setting in Appendix~\ref{app:cribC}, we only give the outline.
As before, we use $T$ transmission blocks to send $T-1$ messages for each user, $(m_1(j))_{j=1}^{T-1}$ and $(m_2(j))_{j=1}^{T-1}$.
Let $\{ p_U p_{X_1|U} , \theta^{x_1} \}$ be a  given ensemble over $\Hset_{A_1}$. Furthermore, consider a measurement instrument $\Wset_{E\to \bar{E}Z}$, and let $\{ p_{X_2|Z,U}(\cdot|z,u) ,  \zeta^{x_2}\}$,  for $z\in\Zset$ and $u\in\Uset$, be a collection of ensembles over $\Hset_{A_2}$.
Define 
\begin{align}
\sigma_{A_1'E}^{x_1}&= \Lset_{A_1\to A_1'E}( \theta^{x_1}) \,,\\
\omega_{A_1' \bar{E}Z}^{x_1}&= \Wset_{E\to \bar{E} Z}(\omega_{A_1'E}^{x_1})\\
\omega_B^{x_1,x_2}&= \channel_{A_1' A_2\to B} (\omega_{A_1'}^{x_1}\otimes \zeta^{x_2})%
\,.
\end{align}
 In addition, for every $u\in\Uset$ and measurement outcome $z\in\Zset$, let
 \begin{align}
\theta^u&=\sum_{x_1\in\Xset_1} p_{X_1|U}(x_1|u) \theta^{x_1}
\\
\zeta^{u,z} &=\sum_{x_2\in\Xset_2} p_{X_2|U,Z}(x_2|u,z) \zeta^{x_2}
\intertext{and consider the corresponding post-measurement states,
}
\omega_{A_1' \bar{E}}^{u,z}&=W_z \Lset_{A_1\to A_1'E}( \theta^{u}) W_z^{\dagger}/ \trace\left(W_z^{\dagger} W_z \Lset_{A_1\to A_1'E}( \theta^{u}) \right)
\\
\omega_B^{u,z}&= \channel_{A_1' A_2\to B} (\omega_{A_1'}^{u,z}\otimes \zeta^{u,z}) \,.
\end{align}
Hence, $\omega_{A_1' \bar{E} Z}$, and $\omega_B$ are the corresponding average states.
The code is %
described below.
\vspace{0.2cm}
\subsection{Classical Codebook Construction}
$\,$
\begin{enumerate}[(i)]
\item
Generate $2^{nR_1}$ independent sequences $u^n(m_0)$, $m_0\in [1:2^{nR_1}]$, at random 
according to $\prod_{i=1}^n p_U(u_{i})$.
\item
For every $m_0$, generate $2^{nR_1}$ conditionally independent sequences $x_1^n(m_0,m_1)$, $m_1\in [1:2^{nR_1}]$, %
according to $\prod_{i=1}^n p_{X_1|U}(x_{1,i}|u_{i}(m_0))$.
\item
For every $m_0$ and measurement sequence $z^n\in\Zset^n$, generate $2^{nR_2}$ conditionally independent sequences $x_2^n(m_0,m_2,z^n)$, $m_2\in [1:2^{nR_2}]$, %
according to $\prod_{i=1}^n p_{X_2|Z,U}(x_{2,i}|z_i,u_{i}(m_0))$.
\end{enumerate}

\subsection{Encoding and Decoding}
\subsubsection*{Encoder 1}
To send the messages $(m_1(j))_{j=1}^{T-1}$,  Alice 1 performs the following.
In block $j$, set  
\begin{align}
&m_0=m_1(j-1) \,,\; m_1=m_1(j) \,.
\intertext{Then, prepare the state}
&\rho_{A_1^n(j)}=\theta^{x_1^n(m_0,m_1)}
\end{align}
 and send $A_1^n(j)$,  for $j\in [1:T]$.
As the $j$th transmission goes through the cribbing channel $\Lset_{A_1\to A_1' E}$, we have 
\begin{align}
\rho_{A_1'^n(j) E^n(j)}=\sigma_{A_1'^n E^n}^{x_1^n(m_0,m_1)} \,.%
\end{align}
The second transmitter can access the cribbing system $E^n(j)$, which are entangled with Alice 1's transmission. 
Cribbing is performed by a sequence of measurements that recover the messages of Alice 1.
In each block, the choice of the cribbing measurement depends on the outcome in the previous block.
With causal encoding, Alice 2 can prepare her input at time $i$ based on the measurement outcome of $E^i$.

\subsubsection*{Encoder 2}
To send the messages $(m_2(j))_{j=1}^{T-1}$,  Alice 2 performs the following. 
Fix $\tm_1(0)\equiv 1$. In block $j$, 
 given the previous cribbing estimate $\tm_1(j-1)$,  do  as follows, for $j=1,2,\ldots, T$.

\begin{enumerate}[(i)]
\item
Apply the measurement instrument $\Wset_{E^n\to  \bar{E}^n Z^n}\equiv \Wset^{\otimes n}_{E\to  \bar{E} Z}$ to the cribbing system $E^n(j)$. As a result, Alice 2 obtains  a measurement outcome $z^n(j)$.
Hence, the average post-measurement state is
\begin{align}
\rho_{A_1'^n(j) \bar{E}^n(j) Z^n(j)}=\omega_{A_1'^n \bar{E}^n Z^n}^{x_1^n(m_0,m_1)} \,.%
\end{align}

\item
Set
\begin{align}
&m_0=\tm_1(j-1)  \,,\; m_2=m_2(j) \,,\; z^n=z^n(j)\,.
\end{align}

\item 
Prepare the state
\begin{align}
&\rho_{A_2^n(j)}=\zeta^{x_2^n(m_0,m_2,z^n)}
\end{align}
 and send $A_2^n(j)$, for $j\in [1:T]$.

\item
 Measure the next cribbing estimate $\tm_1(j)$ by applying a POVM $\Kset_{\bar{E}^n(j) Z^n(j)|m_0}=\{ K_{m_1|m_0} \}$, which will be chosen later, on the joint system $\bar{E}^n(j),Z^n(j)$ following step (i). %
\end{enumerate}

\subsection{Backward decoding}
	The decoder recovers the messages in a backward order, %
	\ie the measurement of the $j$th message of Alice 1 is chosen based on the estimate of $m_1(j+1)$. Fix $\hm_1(0)=\hm_1(T)\equiv 1$.
In block $j$,  for $j=T,T-1,\ldots,2$, the decoder uses the previous estimate of $\hm_1(j)$, and 
  measures $\hm_1(j-1)$ and $\hm_2(j)$ using a POVM 
	$\Dset_{B^n(j)|\hm_1}=\{  D%
	_{m_0,m_2|\hm_1} \}$ with $\hm_1=\hm_1(j)$. Finally, in block $1$, the decoder measures $\hm_2(1)$ with a POVM 
	$\Dset_{B^n(1)|m_0,m_1}=\{  D'%
	_{m_2|\hm_0,\hm_1} \}$, with $\hm_0=1$ and $\hm_1=\hm_1(1)$. The decoding POVMs will also be specified later.

\subsection{Analysis of Probability of Error}
Let $\delta>0$. We use the notation $\eps_i(\delta)$, $i=1,2,\ldots$,
for terms that tend to zero as $\delta\rightarrow 0$.
By symmetry, we may assume without loss of generality that the transmitters send the messages $M_1(j)=M_2(j)=1$.
Consider the following events,
\begin{align}
\mathscr{E}_0(j)=& \{  (U^n(M_1(j-1)),X_1^n(M_1(j-1),1),%
X_2^n(M_1(j-1),1,Z^n(j)))\notin \Aset^{\delta_1}(p_{U,X_1,X_2}) \} \\
\mathscr{E}_1(j)=& \{  \tM_1(j)\neq 1  \}\\
\mathscr{E}_2(j)=& \{  \hM_1(j-1) \neq 1  %
\}
\\
\mathscr{E}_3(j)=& \{  \hM_2(j)\neq 1 \} 
\end{align}
for $j\in [1:T]$, with $\delta_1\equiv \delta/(2 |\Xset_2| |\Uset|)$.
By the union of events bound, the probability of error is bounded by
\begin{align}
P_{e|m_1(j)=m_2(j)=1}^{(Tn)}(\Fset_1,\Fset_2,\Kset,\Dset) %
&\leq %
\sum_{j=1}^{T} \prob{ \mathscr{E}_0(j) }%
+\sum_{j=1}^{T} \cprob{ \mathscr{E}_1(j) }{ \mathscr{E}_0^c(j)\cap \mathscr{E}_1^c(j-1) } \nonumber\\&
+\sum_{j=1}^{T}\cprob{ \mathscr{E}_2(j)\cup \mathscr{E}_3(j) }{ \mathscr{E}_0^c(j)\cap \mathscr{E}_1^c(j) \cap \mathscr{E}_2^c(j+1) } %
\label{eq:PeBc}
\end{align}
where the conditioning on $M_1(j)=M_2(j)=1$ is omitted for convenience of notation.
By the weak law of large numbers, the probability terms $\prob{ \mathscr{E}_0(j) }$ tend to zero as $n\rightarrow\infty$.

To bound the second sum, which is associated with the cribbing measurements, we use the quantum packing lemma.
Alice 2's measurement is effectively a decoder for the marginal cribbing channel 
$\Kset_{E\to\bar{E}Z}\circ\Lset^{(1)}_{A_1\to E}$.
 Given $\mathscr{E}_0^c(j)$, we have that $(U^n(M_1(j-1)),X_1^n(M_1(j-1),1))\in \Aset^{\delta}(p_{U,X_1}) $. %
Now, observe that
\begin{align}
&\trace\left[ \Pi^{\delta}(\omega_{\bar{E}Z}|u^n,x_1^n) \omega_{\bar{E}^n Z^n}^{u^n,x_1^n} \right] \geq 1-\eps_1(\delta) 
\\
&\Pi^{\delta}(\omega_{\bar{E}Z}|u^n)  \omega_{\bar{E}^n Z^n}^{u^n}   \Pi^{\delta}(\omega_{\bar{E}Z}|u^n) \preceq %
2^{ -n(H(\bar{E} Z|U)_{\omega}-\eps_1(\delta)) } \Pi^{\delta}(\omega_{\bar{E} Z})
\\
&\trace\left[ \Pi^{\delta}(\omega_{\bar{E} Z}|u^n,x_1^n)  \right] \leq 2^{ n(H(\bar{E} Z|U,X_1)_{\omega} +\eps_1(\delta))} \\
&\trace\left[ \Pi^{\delta}(\omega_{\bar{E} Z}|u^n) \omega_{\bar{E}^n Z^n}^{u^n,x_1^n} \right] \geq 1-\eps_1(\delta) 
\end{align}
for all $(u^n,x_1^n)\in\Aset^{\delta}(p_{U,X_1})$, by 
(\ref{eq:UnitTCond})-(\ref{eq:UnitTCondB}), respectively.
Since the codebooks are statistically independent of each other, we have by the single-user quantum packing lemma \cite{HsiehDevetakWinter:08p}   \cite[Lemma 12]{Pereg:22p}, that there exists a POVM $K_{m_1|u^n}$ such that
$%
\cprob{ \mathscr{E}_1(j) }{ \mathscr{E}_0^c(j)\cap \mathscr{E}_1^c(j-1) } \leq 2^{ -n( I(X_1;\bar{E}Z|U)_\rho -R_1-\eps_2(\delta)) } 
$, %
which tends to zero as $n\rightarrow\infty$, provided that 
\begin{align}
R_1< I(X_1;\bar{E}Z|U)_\omega -\eps_2(\delta) \,.
\label{eq:Er1C}
\end{align}
The last sum in the RHS of (\ref{eq:PeBc}) tends to zero as in Appendix~\ref{app:cribC}, based on the quantum multi-party packing lemma \cite{DingGharibyanHaydenWalter:20p}, 
for
provided that 
$%
R_2< I(X_2;B|X_1 U)_\omega
$ and %
$R_1+R_2<I(X_1 X_2;B)_\omega $. %
This completes the proof outline for part 1.

\subsection*{Part 2}
Consider the classical-quantum MAC $\channel_{X_1 X_2\to B}\circ \text{id}_{X_1\to X_1 X_1}$ with either causal or non-causal noiseless  cribbing. Observe that it suffices to show the direct part for causal cribbing, and the converse part for the non-causal setting, since $\opC_{\text{caus}}(\channel\circ\Lset)\subseteq \opC_{\text{n-c}}(\channel\circ\Lset)$ (see Remark~\ref{rem:inclusion}).
As the cribbing system stores a perfect copy of Alice 1's classical input in this case, \ie $E\equiv \bar{E}\equiv Z\equiv X_1$, we have $I(X_1;\bar{E} Z|U)_\omega=H(X_1|U)$. Thereby, achievability immediately follows from part 1 of the theorem.

Now, we show the converse part for non-causal cribbing.
 Suppose that Alice $k$ chooses a message $M_k$ uniformly at random, for $k=1,2$. Alice 1 transmits $X_1^n=\Fset_1(M_1)$. Thereby,  Alice 2 measures $X_1^n$  from the cribbing system and transmits $X_2^n=\Fset_2(M_2,X_1^n)$. 
Then, $X_1^n$ and $X_2^n$ are sent through $n$ copies of the classical-quantum MAC $\channel_{X_1 X_2\to B}$. Bob receives the output systems $B^n$ and performs a measurement in order to obtain an estimate $(\hM_1,\hM_2)$ of the message pair.
Consider a sequence of codes %
such that the average probability of error tends to zero.
Since the encoding map $\Fset_1$ is  classical, there exists a random element
$S_1$ such that %
$X_1^n\equiv f_1(M_1,S_1)$, where 
$f_1$ is a deterministic function. 
Then, Alice 1's transmission rate satisfies
\begin{align} 
nR_1
&= %
H(X_1^n|S_1)
\nonumber\\
&\leq %
\sum_{i=1}^n H(X_{1,i}) 
\,.
\label{eq:ConvIneq1Cbcq}
\end{align}

To bound the second transmission rate %
and the rate sum, %
we apply Fano's inequality as in Appendix~\ref{app:cribC}, hence
\begin{align}
nR_2
&\leq I(M_2;B^n|M_1,S_1 )_\rho+n\eps_n '' 
\label{eq:ConvIneq2Cbcq1}
\end{align}
 (see (\ref{eq:ConvIneq2SCbcq1})). Using the chain rule, the last bound can be expressed as
\begin{align}
nR_2&\leq \sum_{i=1}^n I(M_2;B_i|M_1,S_1,B^{i-1} )_\rho+n\eps_n '' 
\nonumber\\
&= \sum_{i=1}^n I(M_2;B_i|M_1,S_1, X_{1,i},  B^{i-1} )_\rho+n\eps_n ''
\label{eq:ConvIneq2Cbcq2} 
\end{align}
 since $X_{1,i}$ is a deterministic function of $(M_1,S_1)$. Observe that %
\begin{align}
 &I(M_2;B_i|M_1,S_1, X_{1,i}, B^{i-1} )_\rho
\nonumber\\
&\leq I(M_2,X_{2,i};B_i|M_1,S_1, X_{1,i}, B^{i-1} )_\rho
\nonumber\\
&= H(B_i|M_1,S_1, X_{1,i}, B^{i-1} )_\rho %
- H(B_i|X_{1,i}, X_{2,i}, M_1,M_2,S_1, B^{i-1} )_\rho
\nonumber\\
&\leq H(B_i |X_{1,i})_\rho
- H(B_i|X_{1,i}, X_{2,i}, M_1,M_2,S_1, B^{i-1} )_\rho \,.
\label{eq:ConvIneq2Cbcq3}
\end{align}
Consider the second term, and observe that %
given $X_{1,i}$ and $X_{2,i}$, the 
system $B_i$ is in a product state with the joint system of $(M_1,M_2,S_1,B^{i-1})$.
Thus, the second term equals $H(B_i| X_{1,i}, X_{2,i} )_\rho$. 
Therefore, by  (\ref{eq:ConvIneq2Cbcq2})-(\ref{eq:ConvIneq2Cbcq3}), 
\begin{align}
nR_2
&\leq 
\sum_{i=1}^n I(X_{2,i};B_i| X_{1,i} )_\rho+\eps_n '' 
\,.
\label{eq:ConvIneq2Cbcq5}
\end{align}
Similarly, the rate sum is bounded by
\begin{align}
n(R_1+R_2)
&\leq 
\sum_{i=1}^n I(X_{1,i} X_{2,i} ;B_i)_{\rho}+n\eps_n ' 
\label{eq:ConvIneq3Cbcq}
\end{align}
(see (\ref{eq:ConvIneq3SCbcq})).

To obtain the single-letter converse, consider a time index $K$ that is drawn uniformly at random, from $[1:n]$, independently of $M_1$, $M_2$, and $S_1$. By (\ref{eq:ConvIneq1Cbcq}), (\ref{eq:ConvIneq2Cbcq5}), and (\ref{eq:ConvIneq3Cbcq}),
\begin{align}
R_1&\leq H(X_{1,K}|K)
\label{eq:ConvIneq1Cbcq2}
\\
R_2-\eps_n '' &\leq I(X_{2,K};B_K| X_{1,K}, K )_\rho
\label{eq:ConvIneq2Cbcq6}
\\
R_1+R_2-\eps_n ' &\leq I(X_{1,K} X_{2,K} ;B_K|K)_{\rho}  \,.
\label{eq:ConvIneq3Cbcq2}
\end{align}
Furthermore, define a joint state $\omega_{X_1 X_2 B}$ by identifying  $X_1\equiv X_{1,K}$, $X_2\equiv X_{2,K}$, and
$B\equiv B_K$. That is,
$%
\omega_{K X_1 X_2 B}\equiv \sum_{i=1}^n \sum_{x_{1},x_{2}} \frac{1}{n} p_{X_{1,i}}(x_{1}) p_{X_{2,i}}(x_{2}) \kb{i}%
\otimes \kb{x_{1}}\otimes \kb{x_{2}}\otimes \channel(x_{1},x_{2}) %
$. %
Thus, the individual rates are bounded by
\begin{align}
R_1&= H(X_1|K)
\nonumber\\
&\leq H(X_1) 
\intertext{and}
R_2-\eps_n '' &\leq I(X_{2};B| X_{1}, K )_\omega
\nonumber\\
&\leq I(X_{2},K;B| X_{1} )_\omega
\nonumber\\
&= I(X_{2};B| X_{1} )_\omega
 \,,
\end{align}
and the rate sum by 
\begin{align}
R_1+R_2-\eps_n ' &\leq I(X_{1} X_{2} ;B|K)_{\omega} 
\nonumber\\
&\leq I(X_{1} X_{2} K ;B)_{\omega} 
\nonumber\\
&= I(X_{1} X_{2} ;B)_{\omega} 
\end{align}
as $I(K;B|X_{1} X_{2})_\omega=0$ since the channel has a memoryless product form.
This completes the proof of Theorem~\ref{theo:cribC}.
\qed

\begin{figure*}

\begin{tabular}{l|ccccc}
{\small Block}				& $1$							& $2$									%
									& $\cdots$			& $T-1$ 		& $T$\\
								
\\ \hline &&&&& \\ 
$U^n$				&	$u^n(1)$	&	$u^n(\ell_1(1))$		%
							& $\cdots$			& $u^n(\ell_1(T-2))$&  $u^n(\ell_1(T-1))$ 
\\&&&&& \\
$X_0^n$				&	$x_0^n(1,\ell_1(1))$	&	$x_0^n(\ell_1(1),\ell_1(2))$		%
							& $\cdots$			& $x_0^n(\ell_1(T-2),\ell_1(T-1))$&  $x_0^n(\ell_1(T-1),1)$ 
\\&&&&& \\
$A_1$				&	$x_1^n(1,\ell_1(1),m_1'(1))$	&	$x_1^n(\ell_1(1),\ell_1(2),m_1'(2))$		%
							& $\cdots$			& $x_1^n(\ell_1(T-2),\ell_1(T-1),m_1'(T-1))$&  $x_1^n(\ell_1(T-1),1,1)$ 
\\&&&&& \\
$E$			& $\widetilde{\ell}_1(1)\rightarrow$				& $\widetilde{\ell}_1(2)\rightarrow$						%
						& $\cdots$			& $\widetilde{\ell}_1(T-1)\rightarrow$ & $\emptyset$
\\&&&&& \\
$A_2$				&	$x_2^n(1,m_2(1))$	&	$x_2^n(\widetilde{\ell}_1(1),m_2(2))$		%
							& $\cdots$			& $x_2^n(\widetilde{\ell}_1(T-2),m_2(T-1))$&  $x_2^n(\widetilde{\ell}_1(T-1),1)$ 
\\&&&&& \\
$B$ 		& $\emptyset$		& $\leftarrow\hat{\ell}_1(1)$ %
					& $\cdots$ & $\leftarrow\hat{\ell}_1(T-2)$ & $\leftarrow\hat{\ell}_1(T-1)$ 
					\\
					& $\hm_2(1)$		& $\hm_2(2)$ %
					& $\cdots$ & $\hm_2(T-1)$ & $\emptyset$
					\\
					& $\hm_1'(1)$		& $\hm_1'(2)$ %
					& $\cdots$ & $\hm_1'(T-1)$ & $\emptyset$
\end{tabular}
\caption[Partial decode-forward cribbing scheme]{Partial decode-forward cribbing scheme. The block index $j\in [1:T]$ is indicated at the top. In the following rows, we have the corresponding elements: 
(1), (2) auxiliary sequences; 
(3) codewords of Alice 1;  
(4) cribbing estimates by Alice 2;
(5) codewords of Alice 2;
(6) estimated messages at the decoder.
The arrows in the fourth row indicate that the Alice 2 measures and encodes forward with respect to the block index, while the arrows in the sixth row indicate that Bob decodes backwards.  
}
\label{fig:PDF}
\end{figure*}

\section{Proof of Theorem~\ref{theo:cribSCp}}
\label{app:cribSCp}
Consider the quantum MAC $\channel\circ\Lset$ with strictly-causal cribbing. 
Here, we prove the partial-decode forward bound on the capacity region. %
We show that for every $\delta_1,\delta_2,\eps_0>0$, there exists a $(2^{n(R_1-\delta_1)},2^{n(R_2-\delta_2)},n,\eps_0)$ code for $\channel_{A_1' A_2\rightarrow B}\circ \Lset_{A_1\to A_1' E}$ with strictly-causal cribbing at Encoder 2, provided that $(R_1,R_2)\in \mathcal{R}^{\text{PDF}}_{\text{s-c}}(\channel\circ\Lset)$. 
To prove achievability, we extend  the classical block Markov coding with backward decoding to the quantum setting, and then apply the quantum packing lemma.

As before, we use $T$ transmission blocks, where each block consists of $n$ channel uses. Given strictly-causal cribbing, Encoder 2 has access to the cribbing measurements from the previous blocks. Alice 1 sends $T-1$  pairs of messages, and Alice 2 sends $T-1$ messages. 
Specifically, Alice 1 sends the pairs  $(\ell_{1}(j),m_{1}'(j))_{j=1}^{T-1}$ at rates $(R_0,R_1-R_0)$, with $R_0<R_1$, %
while Alice 2 sends  $(m_2(j))_{j=1}^{T-1}$ at rate $R_2$.
Let us fix $\ell_{1}(0)=m_{1}'(0)=m_2(0) \equiv 1$ and $\ell_{1}(T)=m_{1}'(T)=m_2(T) \equiv 1$.
 Hence, the coding rate for User $k$ is
$\left(\frac{T-1}{T}\right) R_k$, which tends to $R_k$ as the number of blocks $T$ grows to infinity.

Let $\{ p_U(u)p_{X_0|U} p_{X_1|X_0,U}(x_1|x_0,u) p_{X_2|X_0,U}(x_2|x_0,u), \theta^{x_1}$ $\otimes \zeta^{x_2} \}$ be a  given ensemble over $\Hset_{A_1}\otimes \Hset_{A_2}$. 
Define 
\begin{align}
\omega_{A_1'E}^{x_1}&= \Lset_{A_1\to A_1'E}( \theta^{x_1}) \,,\\
\omega_B^{x_1,x_2}&= \channel_{A_1' A_2\to B} (\omega_{A_1'}^{x_1}\otimes \zeta^{x_2})%
\,.
\end{align}
 In addition, let
\begin{align}
\theta^{x_0,u}&=\sum_{x_1\in\Xset_1} p_{X_1|X_0,U}(x_1|x_0,u) \theta^{x_1}
\\
\zeta^{x_0,u} &=\sum_{x_2\in\Xset_2} p_{X_2|X_0,U}(x_2|x_0,u) \zeta^{x_2}
\\
\omega_{A_1' E}^{x_0,u}&=\Lset_{A_1\to A_1'E}( \theta^{x_0,u})
\\
\omega_B^{x_0,u}&= \channel_{A_1' A_2\to B} (\omega_{A_1'}^{x_0,u}\otimes \zeta^{x_0,u})
\end{align}
for $u\in\Uset$ and $x_0\in\Xset_0$.
Hence, $\omega_{A_1' E}$ and $\omega_B$ are the corresponding average states.

\vspace{0.2cm}
The partial decode-forward coding scheme %
is described below and depicted in Figure~\ref{fig:PDF}.
\vspace{0.2cm}
\subsection{Classical Codebook Construction}
$\,$
\begin{enumerate}[(i)]
\item
Generate $2^{nR_0}$ independent sequences $u^n(\ell_{0})$, $\ell_{0}\in [1:2^{nR_0}]$, at random 
according to $\prod_{i=1}^n p_U(u_{i})$.
\item
For every $\ell_{0}$, generate $2^{nR_0}$ conditionally independent sequences $x_0^n(\ell_0,\ell_1)$, $\ell_1 \in [1:2^{nR_0}]$,  %
according to $\prod_{i=1}^n p_{X_0|U}(x_{0,i}|u_{i}(\ell_0))$.
\item
For every $(\ell_0,\ell_1)$, generate $2^{n (R_{1}-R_0)}$ conditionally independent sequences $x_1^n(\ell_0,\ell_1,m_{1}')$,  $m_{1}'\in [1:2^{n(R_{1}-R_0)}]$, %
according to $\prod_{i=1}^n p_{X_1|U}(x_{1,i}|x_{0,i}(\ell_0,\ell_1),u_{i}(\ell_0))$.
\item
For every $\ell_0$, generate $2^{nR_2}$ conditionally independent sequences $x_2^n(\ell_0,m_2)$, $m_2\in [1:2^{nR_2}]$, %
according to $\prod_{i=1}^n p_{X_2|U}(x_{2,i}|u_{i}(m_0))$.
\end{enumerate}

\subsection{Encoding and Decoding}

\subsubsection*{Encoder 1}
To send the messages $(\ell_{1}(j),m_{1}'(j))_{j=1}^{T-1}$,  Alice 1 performs the following.
In block $j$, set  
\begin{align}
&\ell_0=\ell_{1}(j-1) \,,\; \ell_{1}=\ell_{1}(j) \,,\; m_{1}'=m_{1}'(j) \,.
\intertext{Then, prepare the state}
&\rho_{A_1^n(j)}=\theta^{x_1^n(\ell_0,\ell_{1},m_{1}')}
\end{align}
 and send $A_1^n(j)$,  for $j\in [1:T]$. See the third row in Figure~\ref{fig:PDF}.
As the $j$th transmission goes through the cribbing channel $\Lset_{A_1\to A_1' E}$, we have 
\begin{align}
\rho_{A_1'^n(j) E^n(j)}=\omega_{A_1'^n E^n}^{x_1^n(\ell_0,\ell_{1},m_{1}')} \,.%
\end{align}
The second transmitter can access the cribbing system $E^n(j)$, which are entangled with Alice 1's transmission. 
Cribbing is performed by a sequence of measurements that recover the messages of Alice 1.
In each block, the choice of the cribbing measurement depends on the outcome in the previous block.
With strictly-causal encoding, Alice 2 must perform the cribbing measurement at the end of the block, after she has already sent $A_2^n(j)$.

\subsubsection*{Encoder 2}
To send the messages $(m_2(j))_{j=1}^{T-1}$,  Alice 2 performs the following. 
Fix $\widetilde{\ell}_{1}(0)\equiv 1$. In block $j$, 
 given the previous cribbing estimate $\widetilde{\ell}_{1}(j-1)$,  do  as follows, for $j=1,2,\ldots, T$.

\begin{enumerate}[(i)]
\item
Set
\begin{align}
&\ell_0=\widetilde{\ell}_{1}(j-1) \,,\; m_2=m_2(j) \,.
\end{align}

\item 
Prepare the state
\begin{align}
&\rho_{A_2^n(j)}= \zeta^{x_2^n(\ell_0,m_2)}
\end{align}
 and send $A_2^n(j)$, for $j\in [1:T]$. See the fourth and fifth rows in Figure~\ref{fig:PDF}.

\item
 Measure the next cribbing estimate $\widetilde{\ell}_{1}(j)$ by applying a POVM $\Kset_{E^n(j)|\ell_0}=
\{ K_{\ell_1|u^n} \}$, with $u^n\equiv u^n(\ell_0)$, that will be chosen later. %

\end{enumerate}

Here, the decoder performs two measurements on each output block.
As in \cite{Pereg:20c1,Pereg:22p}, we use the gentle measurement lemma to verify that the decoder's measurement does not cause a ``state collapse" at the output.
\subsubsection*{Sequential decoding}%
To estimate Alice 1 and Alice 2's messages, Bob performs the following. 
\begin{enumerate}[(i)]
\item 
First, the decoder recovers the messages $(\ell_1(j))$ and $(m_2(j))$, using backward decoding.
That is, the measurement of the $j$th message is chosen based on the estimate of $\ell_1(j+1)$. See the bottom part of Figure~\ref{fig:PDF}. Fix $\ell_1(0)=\ell_1(T)\equiv 1$.
In block $j$,  for $j=T,T-1,\ldots,1$, the decoder uses the previous estimate of $\hat{\ell}_{1}(j)$, and 
  measures $\hat{\ell}_{1}(j-1)$ and $\hm_2(j)$ using a POVM 
	$\Dset_{B^n(j)|\ell_{1}}=\{  D%
	_{\ell_0,m_2|\ell_{1}} \}$ with $\ell_{1}=\hat{\ell}_{1}(j)$. 

\item
Next, the decoder recovers the messages $(m_1'(j))$, going in the forward direction. 
For $j=1,2,\ldots,T$, the decoder uses the estimate of $\hat{\ell}_{1}(j-1)$, $\hat{\ell}_{1}(j)$, and $\hm_2(j)$ from the previous step, and 
  measures $\hm_1'(j)$ using a POVM 
	$\Gset_{B^n(j)|\ell_0,\ell_1,m_2}=\{  G%
	_{m_1'|x_0^n,x_2^n} \}$, based on the knowledge of %
		$\ell_{0}=\hat{\ell}_{1}(j-1)$, $\ell_{1}=\hat{\ell}_{1}(j)$, $m_2=\hm_2(j)$, $x_0^n=x_0^n(\ell_0,\ell_1)$, and $x_2^n=x_2^n(\ell_0,m_2)$. 
\end{enumerate}

\subsection{Analysis of Probability of Error}
Let $\delta>0$. We use the notation $\eps_i(\delta)$, $i=1,2,\ldots$,
for terms that tend to zero as $\delta\rightarrow 0$.
By symmetry, we may assume without loss of generality that the transmitters send the messages $L_{1}(j)=M_{1}'(j)=M_2(j)=1$.
Consider the following events,
\begin{align}
\mathscr{E}_0(j)=& \{  (U^n(L_1(j-1)),X_0^n(L_1(j-1),1),%
X_1^n(L_1(j-1),1,1),X_2^n(L_1(j-1),1))%
\notin \Aset^{\delta_1}(p_{U,X_0,X_1,X_2}) \} \\
\mathscr{E}_1(j)=& \{  \widetilde{L}_{1}(j)\neq 1  \}\\
\mathscr{E}_2(j)=& \{  \widehat{L}_{1}(j-1) \neq 1   \}
\\
\mathscr{E}_3(j)=& \{  \hM_2(j))\neq 1 \} 
\\
\mathscr{E}_4(j)=& \{  \hM_{1}'(j) \neq 1 \}
\end{align}
for $j\in [1:T]$, with $\delta_1\equiv \delta/(2 |\Xset_2| |\Uset|)$.
By the union of events bound, the probability of error is bounded by
\begin{align}
&P_{e|m_1(j)=m_2(j)=1}^{(Tn)}(\Fset_1,\Fset_2,\Kset,\Dset) \nonumber\\& \leq %
\sum_{j=1}^{T} \prob{ \mathscr{E}_0(j) }%
+\sum_{j=1}^{T} \cprob{ \mathscr{E}_1(j) }{ \mathscr{E}_0^c(j)\cap \mathscr{E}_1^c(j-1) } \nonumber\\&
+\sum_{j=1}^{T}\cprob{ \mathscr{E}_2(j)\cup \mathscr{E}_3(j) }{ \mathscr{E}_0^c(j)\cap \mathscr{E}_1^c(j) \cap \mathscr{E}_2^c(j+1) } 
\nonumber\\&
+\sum_{j=1}^{T}\Pr\big( \mathscr{E}_4(j) \mid \mathscr{E}_0^c(j)\cap \mathscr{E}_1^c(j) \cap \mathscr{E}_2^c(j+1) %
\cap \mathscr{E}_2^c(j)\cap \mathscr{E}_3^c(j)  \big)
\label{eq:PeBscP}
\end{align}
where the conditioning on $L_{1}(j)=M_{1}'(j)=M_2(j)=1$ is omitted for convenience of notation.
By the weak law of large numbers, the first sum %
tends to zero as $n\rightarrow\infty$.
To bound the second and third sums, we use the arguments in Appendix~\ref{app:cribSCp}, replacing $m_0,m_1$ by $\ell_0,\ell_1$, respectively, and thus, replacing $R_1$ and $X_1$ by $R_0$ and $X_0$, respectively. 
Thus, %
there exists a cribbing measurement $K_{\ell_1|u^n}$ such that  the second sum %
tends to zero if %
\begin{align}
R_{0}< I(X_0;E|U)_\omega -\eps_2(\delta) \,,
\label{eq:achR0}
\end{align}
by the single-user packing lemma  (see (\ref{eq:Er1})). Furthermore,
there exists a POVM $D_{\ell_0,m_2|\ell_1}$ such that the third sum tends to zero as $n\to\infty$,
provided that 
\begin{align}
R_2&< I(X_2;B|X_0 U)_\omega-\eps_4(\delta)
\label{eq:achR2}
\intertext{and}
R_0+R_2&<I(X_0 X_2;B)_\omega-\eps_4(\delta) %
\label{eq:achR0R2}
\end{align}
by the quantum multi-packing lemma \cite{DingGharibyanHaydenWalter:20p} 
and the arguments  in Appendix~\ref{app:cribSCp} that lead to  (\ref{eq:B30})-(\ref{eq:B31}). %

It remains to show that the fourth sum in the RHS of (\ref{eq:PeBscP}) tends to zero as well.
As in \cite{Pereg:20c1,Pereg:22p}, we observe that 
the gentle measurement lemma \cite{Winter:99p,OgawaNagaoka:07p} implies that the post-measurement state $\tilde{\rho}_{B^n}$ is close to the original state $\rho_{B^n}$ in the sense that
\begin{align}
\frac{1}{2}\norm{\tilde{\rho}_{B^n}-\rho_{B^n}}_1 
&\leq \eps_5(\delta)
\end{align}
for sufficiently large $n$ and rates as in (\ref{eq:achR2})-(\ref{eq:achR0R2}).
Therefore, the distribution of measurement outcomes when $\tilde{\rho}_{B^n}$ is measured is roughly the same as if the measurements
$\Dset_{B^n(j)|L_1(j-1)}$ were never performed. To be precise, the difference between the probability of a measurement outcome $\hm_1'$ when $\tilde{\rho}_{B^n}$ is measured and the probability when $\rho_{B^n}$ is measured is bounded by $ \eps_5(\delta)$ in absolute value (see  \cite[Lem. 9.11]{Wilde:17b}).
Furthermore, 
\begin{align}
&\trace\left[ \Pi^{\delta}(\omega_{B}|x_1^n,x_0^n,x_2^n) \omega_{{B_1}^n}^{x_1^n,x_0^n,x_2^n} \right] \geq 1-\eps_6(\delta) \\
&\Pi^{\delta}(\omega_B|x_0^n,x_2^n)  \omega_{{B}^n}^{x_1^n,x_0^n,x_2^n}   \Pi^{\delta}(\omega_{B}|x_0^n,x_2^n) 
\preceq%
2^{ -n(H({B}|X_0 X_2)_{\omega}-\eps_6(\delta)) } \Pi^{\delta}(\omega_{B}|x_0^n,x_2^n)
\\
&\trace\left[ \Pi^{\delta}(\omega_{B}|x_1^n,x_0^n,x_2^n)  \right] \leq 2^{ n(H({B}|X_1 X_0 X_2)_{\omega} +\eps_6(\delta))} \\
&\trace\left[ \Pi^{\delta}(\omega_{B}|x_0^n,x_2^n) \omega_{{B}^n}^{x_1^n,x_0^n,x_2^n} \right] \geq 1-\eps_6(\delta) 
\end{align}
for all $(x_0^n,x_1^n,x_2^n)\in\Aset^{\nicefrac{\delta}{2}}(p_{X_0,X_1,X_2})$, by (\ref{eq:UnitTCond})-(\ref{eq:UnitTCondB}), respectively. 
Thus, by the single-user quantum packing lemma \cite{HsiehDevetakWinter:08p}   \cite[Lemma 12]{Pereg:22p}, there exists a POVM $G_{m_1'|x_0^n,x_2^n}$ such that the error term in the fourth sum 
tends to zero as $n\rightarrow\infty$, provided that 
\begin{align}
R_1-R_0< I(X_1;B|X_0 X_2)_\omega -\eps_2(\delta) \,.
\label{eq:achR1R0}
\end{align}
We have thus shown that a rate pair $(R_1,R_2)$ is achievable if
\begin{align}
R_1&>R_0 \nonumber\\
R_0&< I(X_0;E|U)_\omega \nonumber\\
R_1-R_0&< I(X_1;B|X_0 X_2)_\omega\nonumber\\
R_0+R_2&< I(X_0 X_2;B)_\omega\nonumber\\
R_2&< I(X_2;B|X_0 U)_\omega
\end{align}
(see (\ref{eq:achR0}), (\ref{eq:achR2}), (\ref{eq:achR0R2}) and (\ref{eq:achR1R0})). By eliminating $R_0$, we obtain the following region 
\begin{align}
R_1&< I(X_0;E|U)_\omega+I(X_1;B|X_0 X_2)_\omega\nonumber\\
R_1+R_2&< I(X_0 X_1 X_2;B)_\omega=I( X_1 X_2;B)_\omega\nonumber\\
R_2&< I(X_2;B|X_0 U)_\omega \,.
\end{align}
Then, set $X_0=(U,V)$. This completes the achievability proof for the partial decode-forward inner bound.
\qed

\section{Proof of Theorem~\ref{theo:cribSCcs}}
\label{app:cribSCcs}
\subsection*{Part 1}
Consider the classical-quantum MAC $\channel_{X_1 X_2\to B}\circ Q_{Z|X_1}$ with strictly-causal noisy  cribbing. 
 Suppose that Alice 1 chooses $m_1$ uniformly at random, and prepares an input state $\rho^{m_1}_{ A_1^n}$. 
Consider a sequence of codes $(\Fset_{1n},\Fset_{2n},\Kset_n,\Dset_n)$ such that the average probability of error tends to zero, hence
the error probabilities $\prob{ \hM_1\neq M_1 |M_2 }$, $\prob{ \hM_2\neq M_2 |M_1 }$, and $\prob{ (\hM_1,\hM_2)\neq (M_1,M_2)}$, 
 are bounded by some
$\alpha_n$ which tends to zero as $n\rightarrow \infty$.
By Fano's inequality \cite{CoverThomas:06b}, it follows that%
\begin{align}
H(M_1|\hM_1,M_2) \leq n\eps_n\\
H(M_2|\hM_2,M_1) \leq n\eps_n'\\
H(M_1,M_2|\hM_1,\hM_2) \leq n\eps_n''
\label{eq:AFWC2scCS}
\end{align}
where $\eps_n,\eps_n',\eps_n''$ tend to zero as $n\rightarrow\infty$.
Hence, 
\begin{align}
nR_1&= H(M_1|M_2,S_1,S_2)\nonumber\\& =I(M_1;\hM_1|M_2,S_2)_{\rho}+H(M_1|\hM_1,M_2,S_1,S_2) 
\nonumber\\
&\leq H(M_1|M_2,S_1,S_2)\nonumber\\& =I(M_1;\hM_1|M_2,S_2)_{\rho}+H(M_1|\hM_1,M_2) 
\nonumber\\
&\leq I(M_1;\hM_1|M_2,S_1,S_2)_{\rho}+n\eps_n \nonumber\\
&\leq I(M_1;B^n |M_2,S_1,S_2)_{\rho}+n\eps_n 
\label{eq:ConvIneq1SCcs}
\end{align}
where the last inequality %
follows from the Holevo bound (see Ref. \cite[Theo. 12.1]{NielsenChuang:02b}). Now, 
\begin{align}
I(M_1;B^n|M_2,S_1,S_2 )_{\rho}%
& \leq I(M_1;B^n Z^n|M_2 S_1 S_2)_\rho
\nonumber\\
&=\sum_{i=1}^n I(M_1;B_i Z_i|B^{i-1} Z^{i-1} M_2 S_1 S_2)_\rho
\nonumber\\
&=\sum_{i=1}^n I(M_1 X_{1,i};B_i Z_i|B^{i-1} Z^{i-1} M_2 S_1 S_2 X_{2,i})_\rho
\label{eq:ConvIneq2SCcs}
\end{align}
where the last equality holds since $X_{1,i}$ and $X_{2,i}$ are deterministic functions of $(M_1,S_1)$ and $(M_2,Z^{i-1},S_2)$, respectively. Observe that
\begin{align}
&I(M_1 X_{1,i};B_i Z_i|B^{i-1} Z^{i-1} M_2 S_1 S_2 X_{2,i})_\rho
\nonumber\\
&= H(B_i Z_i|B^{i-1} Z^{i-1} M_2 S_1 S_2 X_{2,i})_\rho%
-H(B_i Z_i|B^{i-1} Z^{i-1} M_1 M_2 S_1 S_2  X_{1,i} X_{2,i})_\rho
\nonumber\\
&\leq  H(B_i Z_i|  X_{2,i})_\rho%
-H(B_i Z_i|B^{i-1} Z^{i-1} M_1 M_2 S_1 S_2  X_{1,i} X_{2,i})_\rho
\label{eq:ConvIneq3SCcs}
\end{align}
since conditioning cannot increase entropy.
The second term equals $H(B_i Z_i|  X_{1,i} X_{2,i})_\rho$ because,
given $(X_{1,i},X_{2,i})=(x_1,x_2)$, the joint cribbing and output system $B_i Z_i$ has no correlation with 
$B^{i-1} Z^{i-1} M_1 M_2 S_1 S_2 $. Thus, by (\ref{eq:ConvIneq1SCcs})-(\ref{eq:ConvIneq3SCcs}),
\begin{align}
R_1&\leq \frac{1}{n}\sum_{i=1}^n I(X_{1,i};B_i Z_i|X_{2,i})_\rho+\eps_n \,.
\end{align}
Similarly,
\begin{align}
R_2%
&\leq \frac{1}{n}\sum_{i=1}^n I(X_{2,i};B_i | X_{1,i})_\rho+\eps_n'
\label{eq:ConvIneq1SCcsR2}
\intertext{and}
R_1+R_2&= %
\frac{1}{n}\sum_{i=1}^n I(X_{1,i} X_{2,i};B_i)_\rho+\eps_n'
\label{eq:ConvIneq3Ccs}
\end{align}

Defining a time index $U$ that is drawn from $[1:n]$ independently uniformly at random, it follows that %
\begin{align}
R_1-\eps_n&\leq I(X_{1};B Z | X_{2} U)_\omega
\label{eq:ConvIneq1Cbcq2cs}
\\
R_2-\eps_n ' &\leq I(X_{2};B| X_{1} U )_\omega
\label{eq:ConvIneq2SCbcq6cs}
\intertext{and}
R_1+R_2-\eps_n '' &\leq I(X_{1} X_{2} ;B|U)_{\omega}
\nonumber\\
&\leq I(X_{1} X_{2} U ;B)_{\omega}
\nonumber\\
&\leq I(X_{1} X_{2} ;B)_{\omega} \,.
\label{eq:ConvIneq3SCbcq2cs}
\end{align}
where $\omega_{UX_1 Z X_2 B}$ is defined as
\begin{align}
&\omega_{UX_1 Z X_2 B}\equiv \frac{1}{n}\sum_{i=1}^n \Big( \kb{i} \otimes\sum_{x_{1},x_{2}}  p_{X_{1,i}}(x_{1}) Q(z|x_1)%
p_{X_{2,i}}(x_{2})  \kb{x_{1},z,x_2}\otimes \channel(x_{1},x_{2}) \Big) \,.
\end{align}

\subsection*{Part 2}
Suppose that $Z=g(X_1)$, where $g:\Xset_1\to\Zset$ is a deterministic function.
We determine the capacity region using the inner bound in Theorem~\ref{theo:cribSCp} and the outer bound in part 1.
Observe that it suffices to consider the first rate, since the first inequality is the only difference between the inner and outer bounds %
(\cf (\ref{eq:inRscP}) and (\ref{eq:inRscCS})).
Consider the partial-decode forward inner bound in Theorem~\ref{theo:cribSCp}. Since $Z=g(X_1)$, we can set $V=Z$ in the RHS of (\ref{eq:inRscP}). Hence, the first rate $R_1$ is bounded by
\begin{align}
I(V;Z|U)=H(Z|U) \,.
\end{align}
The converse part follows from part 1, as
\begin{align}
I(X_{1};B Z | X_{2} U)_\omega&= I(X_{1};Z | X_{2} U)+I(X_{1};B  | X_{2} UZ)_\omega
\nonumber\\
&\leq H(Z | X_{2} U)+I(X_{1};B  | X_{2} UZ)_\omega
\nonumber\\
&\leq H(Z |  U)+I(X_{1};B  | X_{2} UZ)_\omega \,.
\end{align}
This completes the proof of Theorem~\ref{theo:cribSCcs}.
\qed

\end{appendices}

\bibliography{references2}{}

\end{document}